\documentclass[journal]{IEEEtran}

\usepackage{fancyhdr}

\usepackage{graphicx}
\usepackage{amsmath}
\usepackage{amsfonts}
\usepackage{optidef}
\usepackage{multirow}
\usepackage{xcolor}
\usepackage{caption}
\usepackage{subcaption}

\usepackage{hyperref}
\usepackage{comment}
\usepackage{lipsum}
\usepackage{graphicx}
\usepackage[utf8]{inputenc}
\usepackage{enumitem}
\usepackage{hyperref}
\usepackage{amsmath}
\usepackage{amsfonts}
\usepackage{bm}
\usepackage{tabularx}
\usepackage{svg}
\usepackage{algpseudocode}
\usepackage{amsmath}
\usepackage{array}
\usepackage[linesnumbered,noend]{algorithm2e}
\usepackage{fixltx2e}
\usepackage{stfloats}
\usepackage[acronyms,nonumberlist,nopostdot,nomain,nogroupskip,hyperfirst=false]{glossaries}

\newacronym{iot}{IoT}{Internet of Things}
\newacronym{lpwan}{LPWAN}{low-power wide-area network}
\newacronym{sf}{SF}{Spreading Factor}
\newacronym{ml}{ML}{machine learning}
\newacronym{ai}{AI}{Artificial Intelligence}
\newacronym{nbiot}{NB-IoT}{Narrowband IoT}
\newacronym{fl}{FL}{Federated Learning}
\newacronym{arq}{ARQ}{Automatic Repeat Request}
\newacronym{fec}{FEC}{Forward Error Correction}
\newacronym{minlp}{MINLP}{Mixed-Integer Nonlinear Programming}
\newacronym{srn}{SRN}{Standard Radio Node}
\newacronym{sdr}{SDR}{software defined radio}
\newacronym{mchem}{MCHEM}{Massive Channel Emulator}

\newacronym{rf}{RF}{radio frequency}
\newacronym{dsa}{DSA}{dynamic spectrum access}
\newacronym{rfp}{RFP}{radio fingerprinting}
\newacronym{shd}{SHD}{spectrum hole detection}
\newacronym{fml}{FML}{federated machine learning}
\newacronym{5g}{5G}{fifth generation}
\newacronym{mmw}{mmWave}{millimeter wave}
\newacronym{pus}{PUs}{primary users}
\newacronym{sus}{SUs}{secondary users}
\newacronym{drl}{DRL}{deep reinforcement learning}
\newacronym{dl}{DL}{deep learning}
\newacronym{dnn}{DNN}{deep neural networks}
\newacronym{ism}{ISM}{Industrial, Scientific, and Medical}
\newacronym{phy}{PHY}{physical layer}
\newacronym{css}{CSS}{chirp spread spectrum}
\newacronym{crc}{CRC}{cyclic redundancy check}

\newacronym{ap}{AP}{access point}
\newacronym{cfr}{CFR}{channel frequency response}
\newacronym{cir}{CIR}{channel impulse response}
\newacronym{cnn}{CNN}{convolutional neural network}
\newacronym{csi}{CSI}{channel state information}
\newacronym{cv}{CV}{computer vision}
\newacronym{har}{HAR}{human activity recognition}
\newacronym{lan}{LAN}{local-area network}
\newacronym{lstm}{LSTM}{long short-term memory}
\newacronym{mimo}{MIMO}{multiple-input multiple-output}
\newacronym{nic}{NIC}{network interface card}
\newacronym{ofdm}{OFDM}{orthogonal frequency-division multiplexing}
\newacronym{ofdma}{OFDMA}{orthogonal frequency-division multiple access}
\newacronym{siso}{SISO}{single-input single-output}
\newacronym{sta}{STA}{station}
\newacronym{wlan}{WLAN}{wireless local-area network}

\newacronym{bfi}{BFI}{beamforming feedback information}

\newacronym{mum}{MU-MIMO}{multi-user \gls{mimo}}

\newacronym{fsl}{FSL}{few-shot learning}
\newacronym{tl}{TL}{transfer learning}
\newacronym{snr}{SNR}{signal-to-noise ratio}
\newacronym{ndp}{NDP}{null data packet}
\newacronym{svd}{SVD}{singular value decomposition}

\newacronym[plural=\gls{ltf}s,firstplural=long training fields (LTFs)]{ltf}{LTF}{long training field}

\RestyleAlgo{ruled}

\usepackage{pgfplots}
\usepackage{tikz}
\usepackage{pgfplotstable}
\usepackage[utf8]{inputenc}
\pgfplotsset{compat=1.17}
\newlength\fheight
\newlength\fwidth
\usetikzlibrary{plotmarks,patterns,decorations.pathreplacing,backgrounds,calc,arrows,arrows.meta,spy,matrix}
\usepgfplotslibrary{patchplots,groupplots,colorbrewer}
\usepackage{tikzscale}
\usepackage{hyperref}
\usepackage{orcidlink}

\newcommand{\FW}{\texttt{BeamSense}\xspace}

\newcommand{\mum}{\gls{mum}\xspace}

\newcommand{\csi}{\gls{csi}\xspace}
\newcommand{\bfi}{\gls{bfi}\xspace}
\newcommand{\ap}{\gls{ap}\xspace}

\newcommand{\st}{\gls{sta}\xspace}
\newcommand{\sts}{\glspl{sta}\xspace}

\begin{document}

\title{\texttt{BeamSense}: Rethinking Wireless Sensing with MU-MIMO Wi-Fi Beamforming Feedback}

\author{Khandaker Foysal Haque\orcidlink{0000-0003-2791-6863},  \IEEEmembership{Graduate Student Member, IEEE}, Milin Zhang,  \IEEEmembership{Graduate Student Member, IEEE}, Francesca Meneghello\orcidlink{0000-0002-9905-0360},  \IEEEmembership{Member, IEEE}, and Francesco Restuccia\orcidlink{0000-0002-9498-2302},  \IEEEmembership{Senior Member, IEEE} \vspace{-0.5cm}
\IEEEcompsocitemizethanks{
\IEEEcompsocthanksitem{Khandaker Foysal Haque, Milin Zhang and Francesco Restuccia are with the Institute for the Wireless Internet of Things, Northeastern University, United States, e-mail: \{haque.k, zhang.mil, frestuc\}@northeastern.edu.}
\IEEEcompsocthanksitem{F. Meneghello is with the Department of Information Engineering, University of Padova, Italy, e-mail: francesca.meneghello.1@unipd.it.}
}}
\thispagestyle{fancy}



\maketitle

\begin{abstract}
In this paper, we propose \FW, a completely novel approach to implement standard-compliant Wi-Fi sensing applications. Wi-Fi sensing enables game-changing applications in remote healthcare, home entertainment, and home surveillance, among others. However, existing work leverages the manual extraction of \gls{csi} from Wi-Fi chips to classify activities, which is not supported by the Wi-Fi standard and hence requires the usage of specialized equipment. On the contrary, \FW leverages the standard-compliant \gls{bfi} to characterize the propagation environment. Conversely from CSI, the \gls{bfi} (i) can be easily recorded without any firmware modification, and (ii) captures the multiple channels between the access point and the stations, thus providing much better sensitivity. \FW includes a novel cross-domain \gls{fsl} algorithm to handle unseen environments and subjects with few additional data points. We evaluate \FW through an extensive data collection campaign with three subjects performing twenty different activities in three different environments. We show that our \gls{bfi}-based approach achieves about 10\% more accuracy when compared to \gls{csi}-based prior work, while our \gls{fsl} strategy improves accuracy by up to 30\% and 80\% when compared with state-of-the-art cross-domain algorithms. 
\end{abstract}


\IEEEpeerreviewmaketitle

\glsresetall

\section{Introduction}
\IEEEPARstart{S}{ince} 1990, Wi-Fi has become the technology of choice for Internet connectivity in indoor environments \cite{WiFiAlliance}. Beyond connectivity, Wi-Fi signals can be used as sounding waveforms to perform activity recognition~\cite{ma2021location}, health monitoring~\cite{wang2017tensorbeat}, and human presence detection~\cite{zhu2017r}, among others~\cite{ma2019wifi}. The intuition behind Wi-Fi sensing is that humans act as obstacles to the propagation of radio signals in the environment. Specifically, when encountering the human body, the radio waves undergo reflections, diffractions and scattering that make the signals collected at the Wi-Fi receiver differ from the transmitted ones. Wi-Fi sensing aims at detecting the changes in the Wi-Fi signals and associating them to the way the subject stays/moves in the environment, thus realizing device-free monitoring solutions. To date, the vast majority of Wi-Fi sensing systems -- discussed in Section \ref{sec:rw} -- leverage channel measurements obtained from pilot symbols as sensing primitive. Such measurements are usually referred to as \gls{csi} and describe the way the signals propagate in the environment. Despite leading to good performance, \gls{csi}-based techniques require extracting and recording the \gls{csi} estimated by the Wi-Fi devices involved in the sensing activities, and such operations are currently not supported by the IEEE 802.11 standard. This has led to the introduction of custom-tailored firmware modifications to extract the \gls{csi}~\cite{csitool2011,athcsi2015,nexmoncsi2019,picoscenes2021,axcsi2021}, which makes the sensing process not scalable. Such \gls{csi} extraction tools only provide support for single-user \gls{mimo} sensing as the channel is sounded on the link between the transmitter and the device implementing the extraction tool. Therefore, Wi-Fi sensing approaches relying on \gls{csi} extraction tools cannot benefit from the spatial diversity that can be gained through \gls{mum} transmissions. Spatial diversity may be achieved considering multiple \gls{csi} collectors but this would increase the computation burden as synchronization among the devices would be needed.
Moreover, even if \gls{csi} extraction could be supported in the future without the need for custom-tailored firmware modifications, it would require additional processing to extract the data from the chip, thus increasing energy consumption. Therefore, we argue that more suitable approaches to Wi-Fi sensing should be put forward.\vspace{-0.4cm}

\begin{figure}[h]
	\centering
	\includegraphics[width=.48\textwidth, height=.27\textwidth]{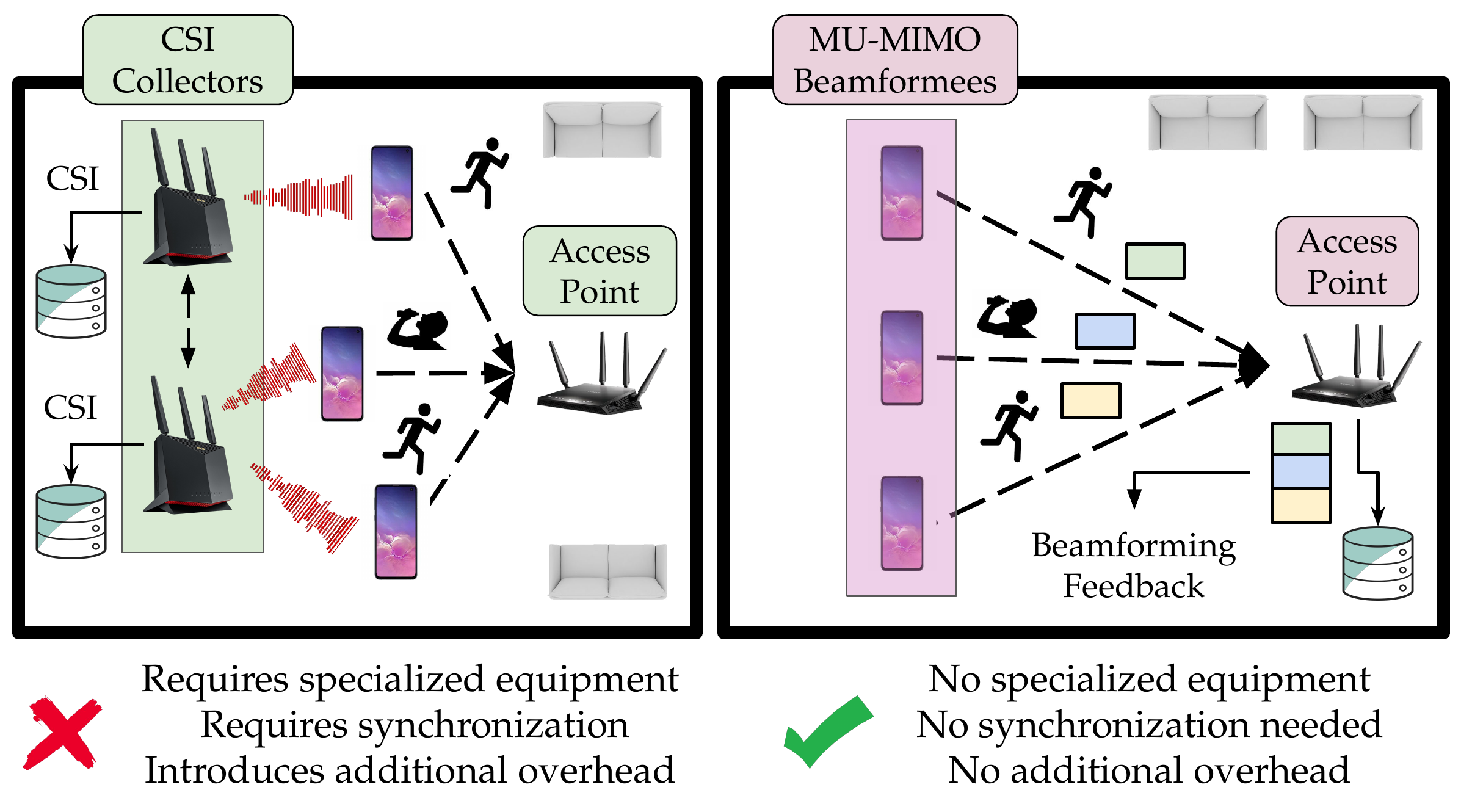}
	\caption{CSI-based vs BFI-based Wi-Fi sensing.}
	\label{fig:csi-vs-bfm}
\end{figure}

In this paper, we propose \FW, an entirely new approach to Wi-Fi sensing that leverages the \mum capabilities of Wi-Fi to drastically increase sensing performance while substantially reducing sensing overhead. As shown in Figure \ref{fig:csi-vs-bfm}, \FW leverages the \gls{bfi} -- traditionally used to beamform transmissions -- to estimate the propagation environment between the \ap and the connected \sts. In stark contrast with \csi-based sensing, \FW (i) does not need firmware modifications, since any off-the-shelf Wi-Fi device can capture \bfi packets, which are sent unencrypted to keep the processing delay below a few milliseconds~\cite{Aryafar2010design}; and (ii) does not require synchronization among receivers, since a single \bfi report contains the information about all the \gls{mimo} channels established between the \ap and the \sts. In fact, while devices empowered with \gls{csi} extraction tools allow obtaining information on a single \gls{mimo} channel, when capturing the \bfi we obtain the channel information associated with all the \sts involved in a \gls{mum} transmission. Thus, multiple spatially diverse channel information is collected with a single capture. 
For this reason, \FW exhibits far better performance in challenging environments, as shown in Section~\ref{sec:per_eva}.\vspace{0.1cm}

\textbf{This paper provides the following contributions:}

\noindent$\bullet$ We propose \FW, a new approach to Wi-Fi sensing where the standard-compliant \gls{bfi} routinely sent in \mum Wi-Fi networks is used to characterize the propagation environment between the \mum users and the \ap. To the best of our knowledge, this is the first work proposing the utilization of \bfi to perform Wi-Fi sensing;

\noindent$\bullet$ We propose a \gls{dl}-based Fast and Adaptive Micro Reptile Sensing (FAMReS) algorithm to perform activity classification based on \bfi. We chose \gls{dl} since it has shown remarkable performance in classifying activities in Wi-Fi sensing settings \cite{bahadori2022rewis}. However, it is well-known that \gls{dl} models may perform poorly when tested in different settings \cite{meneghello2022sharp}. For this reason, FAMReS leverages \gls{fsl} to quickly generalize to different subjects and environments with few additional data points;

\noindent$\bullet$ We extensively evaluate \FW through a comprehensive data collection campaign, with three subjects performing twenty different activities in three different environments. For that, we built a reconfigurable IEEE 802.11ac MU-MIMO network with three \glspl{sta} and one \gls{ap}. The Wi-Fi network was also synchronized with a camera-based system that records the ground truth for our experiments and a secondary IEEE 802.11ac network empowered with Nexmon CSI~\cite{nexmoncsi2019} to concurrently collect the \gls{csi} measurements used for comparative analysis.
We show that our BFI-based approach combined with a traditional convolutional neural network (CNN) without pre-processing achieves about 10\% more accuracy when compared to state-of-the-art \gls{csi}-based techniques, which uses pre-processing. Moreover, FAMReS improves accuracy by up to 30\% and 80\% when compared with state-of-the-art cross-domain algorithms. \textbf{For reproducibility, we will release the entirety of our 800 GB dataset and our code.}

The rest of the article is organized as follows. In Section~\ref{sec:rw} we review the existing literature in the area. The \FW Wi-Fi sensing system is illustrated in Section~\ref{sec:BeamSense} whereas the performance evaluation of the system is presented in Section~\ref{sec:per_eva}. Section~\ref{sec:conclusion} concludes the discussion. 

\thispagestyle{fancy} 

\section{Related Work}\label{sec:rw}

Over the last ten years, a lot of efforts have been made to explore wireless sensing, which is summarized by Liu et al. in \cite{liu2019wireless}. The first Wi-Fi sensing approaches were based on the received signal strength indicator (RSSI) \cite{hsieh2020device,wang2015understanding,zhang2022domain,depatla2018crowd,ssekidde2021augmented,singh2021machine}. More recently, researchers have focused on the more fine-grained \gls{csi} information that describes how the wireless channel modifies signals at different frequencies rather than providing a cumulative metric on the signal attenuation as the RSSI does. Passive Wi-Fi radar (PWR)-based approaches~\cite{li2020taxonomy,li2020passive,tang2022people,tang2020occupancy,huang2019pedestrian} have also been proposed in the literature. However, such an approach requires specialized hardware (\gls{sdr}) to analyze the collected signal. In the rest of the section, we focus on \gls{csi}-based sensing, and summarize the main research on the topic.\vspace{0.1cm}

\noindent\textbf{Background on CSI-based Sensing.}~The term \gls{csi} can refer both to the time-domain \gls{cir} or the frequency-domain \gls{cfr}. Specifically, the \gls{cir} encodes the information about the multipath propagation of the transmitted signal: each peak in the \gls{cir} represents a propagation path characterized by a specific time delay (linked with the length of the path) and an attenuation. Multipath propagation is a typical phenomenon of indoor environments, where obstacles (objects, people, animals) in the surroundings act as reflectors/diffractors/scatterers for the irradiated wireless signals. In turn, the receiver collected different copies of the transmitted signal each associated with a different propagation, or, equivalently, an obstacle in the environment. The \gls{cfr} represents the Fourier transform of the \gls{cir} and describes how the environment modifies signals transmitted with different carrier frequencies. 
Specifically, indicating with $\mathbf{x}(f,t)$ and $\mathbf{y}(f,t)$ the frequency domain representation of the transmitted and received signals at time $t$ and frequency $f$ respectively, and with $\mathbf{h}(f,t)$ the \gls{cfr}, we have that $\mathbf{y}(f,t) = \mathbf{h}(f,t) \times \mathbf{x}(f,t)$~\cite{bu2022transfersense}. Considering the $M\times N$ \gls{mimo} \gls{ofdm} system, with $K$ sub-channels, and $M$ and $N$ transmitting and receiving antennas respectively, the \gls{cfr} is a $K\times M \times N$-dimensional matrix providing the amplitude and phase information over each \gls{ofdm} sub-channel for any given pair of transmitting and receiving antenna.\vspace{0.1cm}


\noindent\textbf{Existing Research on CSI-based Sensing.}~Over the last decade, CSI-based sensing has been proposed for a wide variety of applications. Among the most compelling, we mention person detection and identification \cite{korany2021,zeng2016wiwho,soltanaghaei2020human}, crowd counting \cite{liu2019deepcount,depatla2018crowd}, respiration monitoring \cite{zeng2020multisense}, baggage tracking \cite{shi2021environment}, smart homes \cite{ren2020liquid,he2020wifi}, human pose tracking \cite{ren2021winect,ren2021tracking,10.1145/3372224.3380900,zhao2018through}, patient monitoring \cite{muaaz2022wi,ge2022contactless}, with most of the previous research activities focusing on human activity recognition (HAR) and human gesture recognition (HGR) \cite{korany2020teaching,wei2019real,zheng2019zero,xiao2021onefi,meneghello2022sharp,ahmed2020device}. \textbf{\textit{The above list is definitely not exhaustive.}} For excellent survey papers on the topic, we refer the reader to \cite{khalili2020wi,ma2019wifi,ma2021location,nirmal2021deep}. In the following, we just summarize the most recent approaches that are most related to the work conducted in this article.
Guo et al. presented WiAR \cite{guo2019wiar}, a CSI-based system achieving up to 90\% accuracy in the recognition of 16 human activities. Similarly, a meta-learning-based approach called RF-Net was presented in \cite{rfnet2020} based on the usage of recurrent neural networks with long short-term memory (LSTM) cells. However, only six activities were considered in the evaluation. Regarding HGR, \cite{zheng2019zero} and \cite{xiao2021onefi} presented Widar 3.0 and OneFi, respectively considering six and forty gestures. The authors in \cite{zheng2019zero} proposed to use a body velocity profile (BVP) measure which has been shown to improve the generalization capability of the classification algorithm. The authors of \cite{xiao2021onefi} used one-shot learning to classify unseen gestures with few labeled samples. The majority of previous work has been evaluated on 802.11n channel data while, to the best of our knowledge, only two works considered HAR in the context of 802.11ac \cite{meneghello2022sharp, bahadori2022rewis}. Meneghello et al. proposed to use the Doppler shift estimated through the CSI to obtain an algorithm that generalizes to different environments \cite{meneghello2022sharp}. Bahadori et al. use instead few-shot learning to achieve environmental robustness \cite{bahadori2022rewis}.\vspace{0.1cm}


\noindent\textbf{Limitations of CSI-based Sensing.}~Since the CSI is computed at the \gls{phy}, it is not readily available with off-the-shelf \glspl{nic}. Although CSI can be extracted with \gls{sdr} implementations, which only support up to 40 MHz of bandwidth, being only IEEE 802.11 a/g/p/n compliant \cite{bloessl2013ieee, bahadori2022rewis}. Moreover, \glspl{sdr} are costly specialized hardware that may be unavailable in real-life situations and require expert knowledge to be used. To overcome such limitations, in recent years, researchers have developed some \gls{csi} extraction tools that run on commercial Wi-Fi \glspl{nic}. Two of them, namely Linux CSI~\cite{csitool2011} and Atheros CSI~\cite{athcsi2015}, target IEEE~802.11n compliant \glspl{nic} (up to 40 MHz bandwidth). The third one, Nexmon CSI~\cite{nexmoncsi2019}, allows extracting the CFR from some IEEE~802.11ac compliant devices, supporting bandwidths up to 80 MHz. The most recent one, AX CSI~\cite{axcsi2021} is designed for IEEE 802.11ax devices and provides \gls{cfr} measurements also on 160 MHz bandwidth channels. These tools, however, need non-trivial firmware modifications of the \glspl{nic}. Moreover, they do not provide support for estimating the channel on \gls{mum} channels. Both when the \gls{csi} extractor tool is implemented on one receiving Wi-Fi device or on another monitor device, only the \gls{mimo} links between the transmitter and the \gls{csi} collector is monitored, i.e., only SU-\gls{mimo} mode is supported. This is a limitation of \gls{csi}-based systems as \gls{mum} systems can provide way richer information than SU-\gls{mimo} ones as they capture the correlation of the propagated signal from different \glspl{sta} relative to the sensed subject. As a last consideration, Wang et al.~\cite{wang2022placement} recently pointed out the importance of the placement of the \gls{csi} extractor device. Specifically, they showed that accurate placement of the sensing devices can enhance the sensing coverage by mitigating severe interference. Non-calibrated placement of the sensing devices can severely hamper the sensing quality.\vspace{0.1cm}

\noindent\textbf{Advantages of \FW.}~Our approach addresses these challenges by exploiting the \gls{mum} beamforming feedback to sense the environment. The collection of the \gls{mum} beamforming feedback packets can be done with any standard-compliant 802.11 ac/ax device, and it does not need any close proximity or direct access to the sensed subject. As our system does not need any specific hardware or infrastructure, it facilitates mass deployment. Moreover, since it utilizes the aggregated feedback from different users placed at different locations, \FW is less sensitive to the accurate placement of the \glspl{sta}.

\section{The \texttt{B\lowercase{eam}S\lowercase{ense}} W\lowercase{i}-F\lowercase{i} Sensing System}\label{sec:BeamSense}

\begin{figure}[t]
	\centering
	\includegraphics[width=\columnwidth]{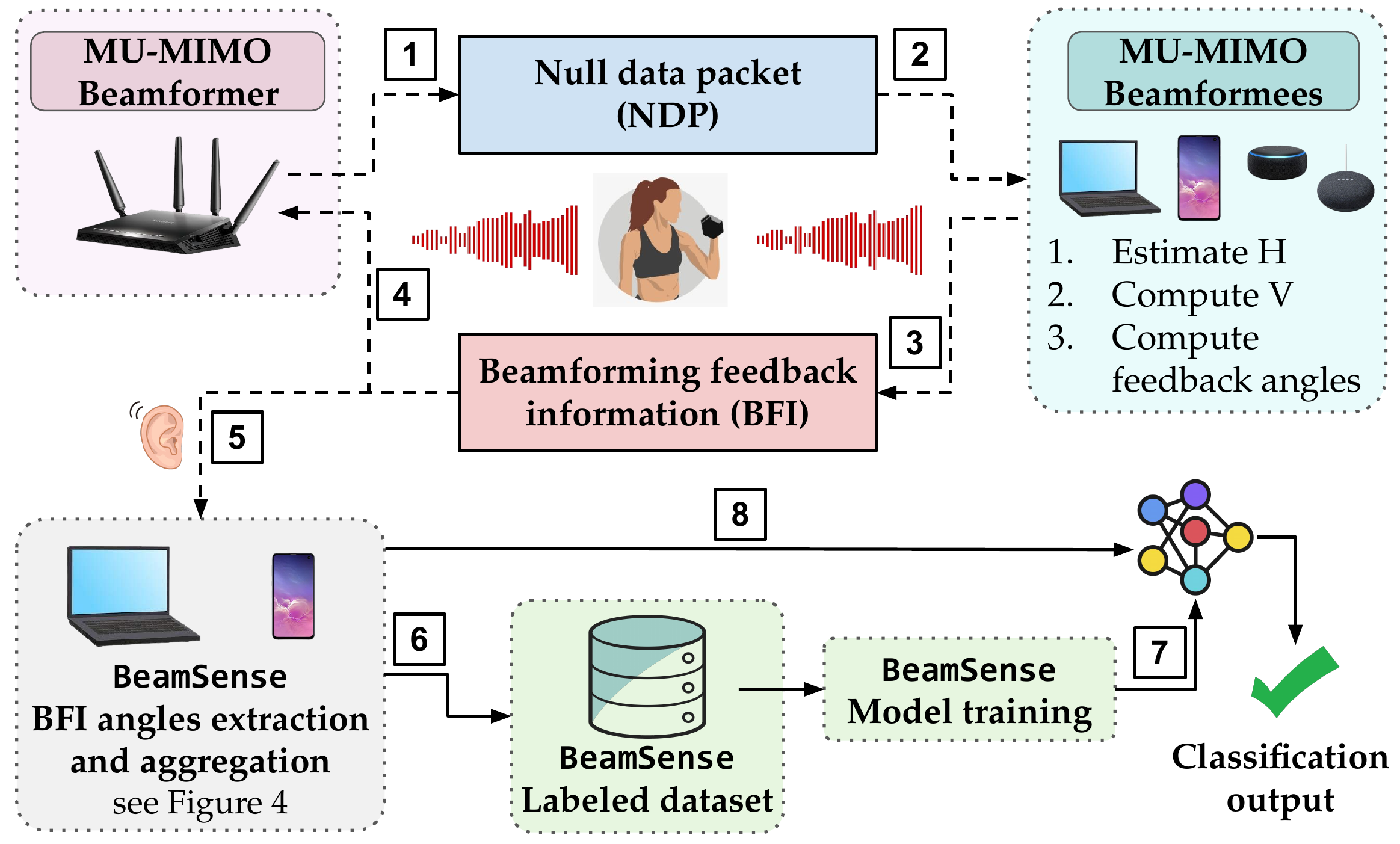}
        \setlength\abovecaptionskip{-0.05cm}
	\caption{The \FW Wi-Fi sensing system.\vspace{-0.5cm}}
	\label{fig:beamsense-fw}
\end{figure}

Figure \ref{fig:beamsense-fw} shows a high-level overview of \FW, which leverages the channel estimation mechanism standardized in IEEE 802.11 to sound the physical environment. The channel estimation is performed on the \sts (beamformees) and is reported to the \ap (beamformer) that uses it to properly beamform \mum transmissions. The report is referred to as the \bfi and is transmitted over the air in clear text. Since the \ap continuously triggers the channel estimation procedure on the connected \sts, \textit{the \bfi contains very rich, reliable, and spatially diverse information}. Moreover, the \bfi \textit{can be collected with a single capture} by the \ap or any other Wi-Fi-compliant device, thus reducing the system complexity.

\smallskip
\noindent\textbf{\FW Technical Challenges.} \FW is a completely novel way to perform Wi-Fi sensing. While previous work in the literature deal with the well-known \gls{csi} data, we instead consider the \gls{bfi} as a sensing primitive. We stress that \gls{bfi} represents a completely new type of data. While \gls{csi} consists of complex I/Q-values, \gls{bfi} is expressed in terms of compressed rotational matrices. In this respect, the first challenge we need to address is the design and implementation of a novel tool to extract the \gls{bfi} data embedded within Wi-Fi frames transmitted from the beamformees to the beamformer as part of the channel sounding procedure. On top of that, the second challenge concerns the implementation of a new data processing pipeline for the new data type that effectively performs activity classification based on \gls{bfi} data and provides environment adaptation features. 
The third challenge to be addressed is the setup of an extensive experimental testbed to implement and assess the performance of the new Wi-Fi sensing approach in a real-world scenario with commercial Wi-Fi devices. 


\smallskip
In the following, we thoroughly detail the \FW sensing system. We use the superscripts $T$ and $\dag$ to denote the transpose and the complex conjugate transpose (i.e., the Hermitian). We define with $\angle{\mathbf{C}}$ the matrix containing the phases of the complex-valued matrix $\mathbf{C}$. Moreover, diag$(c_1, \dots, c_j)$ indicates the diagonal matrix with elements $(c_1, \dots, c_j)$ on the main diagonal. The $(c_1, c_2)$ entry of matrix $\mathbf{C}$ is defined by $\left[\mathbf{\mathbf{C}}\right]_{c_1, c_2}$, while $\mathbb{I}_{c}$ refers to an identity matrix of size $c \times c$ and $\mathbb{I}_{c\times d}$ is a $c \times d$ generalized identity matrix.

\subsection{\FW: A Walkthrough}\label{sec:mu-mimo}

\begin{figure}[t]
	\centering
	\includegraphics[width=.37\textwidth, height=.29\textwidth]{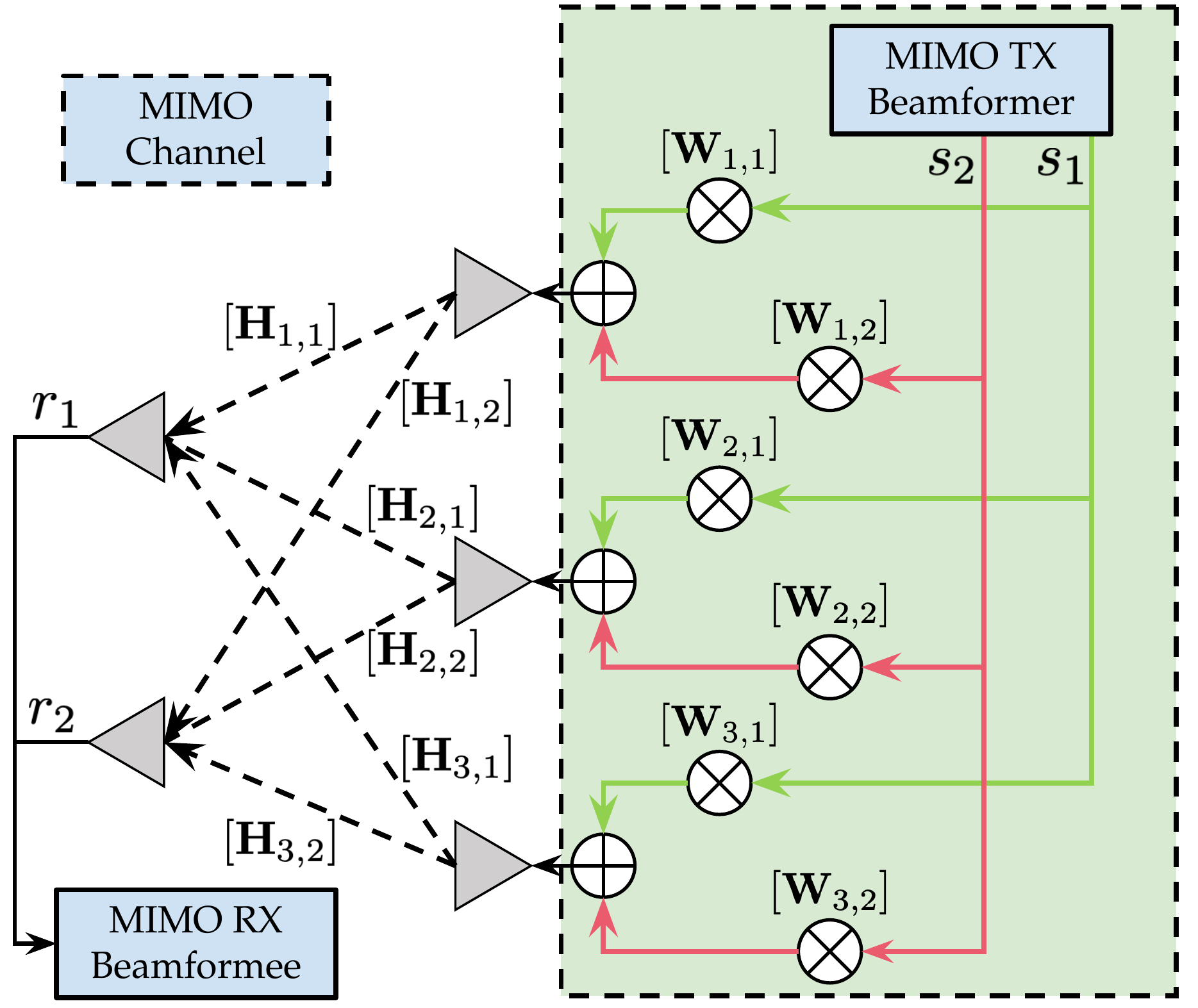}
	\caption{Example of $3 \times 2$ MIMO system. ${s_1, s_2}$ and ${r_1, r_2}$ are respectively the transmitted and received signals. The symbol $\mathbf{W}$ indicates the steering matrix, while $\mathbf{H}$ is the \gls{cfr}.\vspace{-0.1cm}}
	\label{fig:mimo-system}
\end{figure}

The \FW sensing system entails eight steps, as depicted in Figure~\ref{fig:beamsense-fw}. The process stems from the way beamforming is implemented in IEEE 802.11 networks. Specifically, the beamformer (\ap) uses a matrix $\mathbf{W}$ of pre-coding weights -- called steering matrix -- to linearly combine the signals to be simultaneously transmitted to the different beamformees (\sts). The steering matrix is derived from the \gls{cfr} matrices $\mathbf{H}$ estimated by each of the beamformee and that describe how the environment modifies the irradiated signals in their path to the receivers. The estimation process is called \textit{channel sounding} and is triggered by the \ap which periodically broadcasts a \gls{ndp} (\textbf{step 1} in Figure \ref{fig:beamsense-fw}) that contains sequences of bits -- named \glspl{ltf} -- the decoded version of which is known by the beamformees. Since its purpose is to sound the channel, the \gls{ndp} \textit{is not beamformed} by the \gls{ap}. \textit{This is particularly advantageous for sensing purposes}, since the resulting \gls{cfr} estimation will not be affected by inter-stream or inter-user interference.
The \glspl{ltf} are transmitted over the different beamformer antennas in subsequent time slots, thus allowing each beamformee to estimate the \gls{cfr} of the links between its receiving antennas and the beamformer transmitting antennas. The \glspl{ltf} are modulated -- as the data fields -- through \gls{ofdm} by dividing the signal bandwidth into $K$ partially overlapping and orthogonal sub-channels spaced apart by $1/T$. The input bits are grouped into \gls{ofdm} symbols, $\mathbf{a} = [a_{-K/2}, \dots, a_{K/2-1}]$, where $a_k$ is named \gls{ofdm} sample. These $K$ \gls{ofdm} samples are digitally modulated and transmitted through the $K$ \gls{ofdm} sub-channels in a parallel fashion thus occupying the channel for $T$ seconds. The transmitted \gls{ltf} signal is
\begin{equation}\label{eq:tx_signal}
	 s_{\rm tx}(t) = e^{j2\pi f_c t} \sum_{k=-K/2}^{K/2-1} a_{k} e^{j2\pi kt/T},
\end{equation}
where $f_c$ is the carrier frequency. 
The \gls{ndp} is received and decoded by each \st (\textbf{step 2}) to estimate the \gls{cfr} $\mathbf{H}$. The different \glspl{ltf} are used to estimate the channel over each pair of transmitting (TX) and receiving (RX) antennas, for every \gls{ofdm} sub-channel. This generates a $K \times M \times N$ matrix $\mathbf{H}$ for each beamformee, where $M$ and $N$ are respectively the numbers of TX and RX antennas. We refer the reader to Section~\ref{sec:rw} for additional details about the \gls{cfr}. Next, the \gls{cfr} is compressed -- to reduce the channel overhead -- and fed back to the beamformer. Using $\mathbf{H}_k$ to identify the $M \times N$ sub-matrix of $\mathbf{H}$ containing the \gls{cfr} samples related to sub-channel $k$, the \textit{compressed beamforming feedback} is obtained as follows (\cite{perahia_stacey_2008}, Chapter~13). First, $\mathbf{H}_k$ is decomposed through \gls{svd} as
\begin{equation}
    \mathbf{H}_k^T = \mathbf{U}_k\mathbf{S}_k\mathbf{Z}_k^\dag,
\end{equation}
where $\textbf{U}_k$ and $\mathbf{Z}_k$ are, respectively, $N \times N$ and $M \times M$ unitary matrices, while the singular values are collected in the $N\times M$ diagonal matrix $\mathbf{S}_k$. Using this decomposition, the complex-valued beamforming matrix $\mathbf{V}_k$ is defined by collecting the first $N_{\rm SS} \le N$ columns of $\mathbf{Z}_k$. Such a matrix is used by the beamformer to compute the pre-coding weights for the $N_{\rm SS}$ spatial streams directed to the beamformee. Hence, $\mathbf{V}_k$ is converted into polar coordinates as detailed in Algorithm~\ref{alg:beamf_feedback} to avoid transmitting the complete matrix. The output is matrices $\mathbf{D}_{k,i}$ and $\mathbf{G}_{k,\ell,i}$, defined as
\begin{equation}
    \mathbf{D}_{k,i} =
	 \begin{bmatrix}
	\mathbb{I}_{i-1} & 0 & \multicolumn{2}{c}{\dots} & 0 \\
	0 & e^{j\phi_{k,i,i}} & 0 & \dots & \multirow{2}{*}{\vdots} \\
	\multirow{2}{*}{\vdots} & 0 & \ddots & 0 &  \\
	 & \vdots & 0 & e^{j\phi_{k,M-1,i}} & 0 \\
	0 & \multicolumn{2}{c}{\dots} & 0 & 1
	 \end{bmatrix},\label{eq:d_matrix}
\end{equation}
\begin{equation}
    \mathbf{G}_{k,\ell,i} =
	 \begin{bmatrix}
	\mathbb{I}_{i-1} & 0 & \multicolumn{2}{c}{\dots} & 0 \\
	0 & \cos{\psi_{k,\ell,i}} & 0 & \sin{\psi_{k,\ell,i}} & \multirow{2}{*}{\vdots} \\
	\multirow{2}{*}{\vdots} & 0 & \mathbb{I}_{\ell-i-1} & 0 &  \\
	 & -\sin{\psi_{k,\ell,i}} & 0 & \cos{\psi_{k,\ell,i}} & 0 \\
	0 & \multicolumn{2}{c}{\dots} & 0 & \mathbb{I}_{M-\ell}
	 \end{bmatrix},\label{eq:g_matrix}
\end{equation}
that allow rewriting $\mathbf{V}_k$ as $\mathbf{V}_k = \mathbf{\Tilde{V}}_k \mathbf{\Tilde{D}}_k$, with
\begin{equation}
    \mathbf{\Tilde{V}}_k = \prod_{i=1}^{\min(N_{\rm SS}, M-1)} \Bigg( \mathbf{D}_{k,i} \prod_{l=i+1}^{M}\mathbf{G}_{k,l,i}^T\Bigg) \mathbb{I}_{M\times N_{\rm SS}}, \label{eq:v_matrix}
\end{equation}
where the products represent matrix multiplications. In the $\mathbf{\Tilde{V}}_k$ matrix, the last row -- i.e., the feedback for the $M$-th transmitting antenna -- consists of non-negative real numbers by construction. Using this transformation, the beamformee is only required to transmit the $\phi$ and $\psi$ angles to the beamformer as they allow reconstructing $\mathbf{\Tilde{V}}_k$ precisely. Moreover, it has been proved (see \cite{perahia_stacey_2008}, Chapter~13) that the beamforming performance is equivalent at the beamformee when using $\mathbf{V}_k$ or $\mathbf{\Tilde{V}}_k$ to construct the steering matrix $\mathbf{W}$. In turn, the feedback for $\mathbf{\Tilde{D}}_k$ is not fed back to the beamformer. The angles are quantized using $b_{\phi} \in \{7, 9\}$ bits for $\phi$ and $b_{\psi} = b_{\phi}-2$ bits for $\psi$, to further reduce the channel occupancy. The quantized values -- \mbox{$q_{\phi} = \{0, \dots, 2^{b_{\phi}}-1\}$} and \mbox{$q_{\psi} = \{0, \dots, 2^{b_{\psi}}-1\}$} -- are packed into the compressed beamforming frame (\textbf{step 3}) and such \textit{beamforming feedback information} (BFI) is transmitted to the \ap (\textbf{step 4}) in \textit{clear text}. Each \bfi contains $A$ number of angles for each of the $K$ \gls{ofdm} sub-channels for a total of $(K \cdot A)$ angles each. In Figure~\ref{fig:mimo-system}, we show an example of how beamforming is conducted in a $3 \times 2$ MIMO system.

\RestyleAlgo{ruled}
\SetKwComment{Comment}{/*}{*/}
\SetAlgoNoLine
\LinesNotNumbered
\begin{algorithm}[t]
\caption{$\mathbf{V}_k$ matrix decomposition}\label{alg:beamf_feedback}
Require: $\mathbf{V}_k$\;
$\mathbf{\Tilde{D}}_k = {\rm diag}(e^{j \angle \left[\mathbf{V}_k\right]_{M,1}}, \dots, e^{j \angle \left[\mathbf{V}_k\right]_{M,N_{\rm SS}}})$ \;
$\mathbf{\Omega}_k = \mathbf{V}_k\mathbf{\Tilde{D}}_k^\dag$\;
\For{$i \leftarrow 1$ to $\min (N_{\rm SS}, M-1)$}{
$\phi_{k,\ell,i} = \angle \left[\mathbf{\Omega}_k\right]_{\ell, i}$ with $\ell={i, \dots, M-1}$\;
compute $\mathbf{D}_{k,i}$ through Eq.~(\ref{eq:d_matrix})\;
$\mathbf{\Omega}_k \leftarrow \mathbf{D}_{k,i}^\dag \mathbf{\Omega}_k$\;
\For{$\ell \leftarrow i+1$ to $M$}{
$\psi_{k,\ell,i} = \arccos \left( \frac{[\mathbf{\Omega}_k]_{i, i}}{\sqrt{[\mathbf{\Omega}_k]_{i, i}^2 + [\mathbf{\Omega}_k]_{\ell, i}^2}} \right)$\;
compute $\mathbf{G}_{k,\ell,i}$ through Eq.~(\ref{eq:g_matrix})\;
$\mathbf{\Omega}_k \leftarrow \mathbf{G}_{k,\ell,i} \mathbf{\Omega}_k$\;}}
\end{algorithm}

\begin{figure}[t]
	\centering
	\includegraphics[width=0.93\columnwidth]{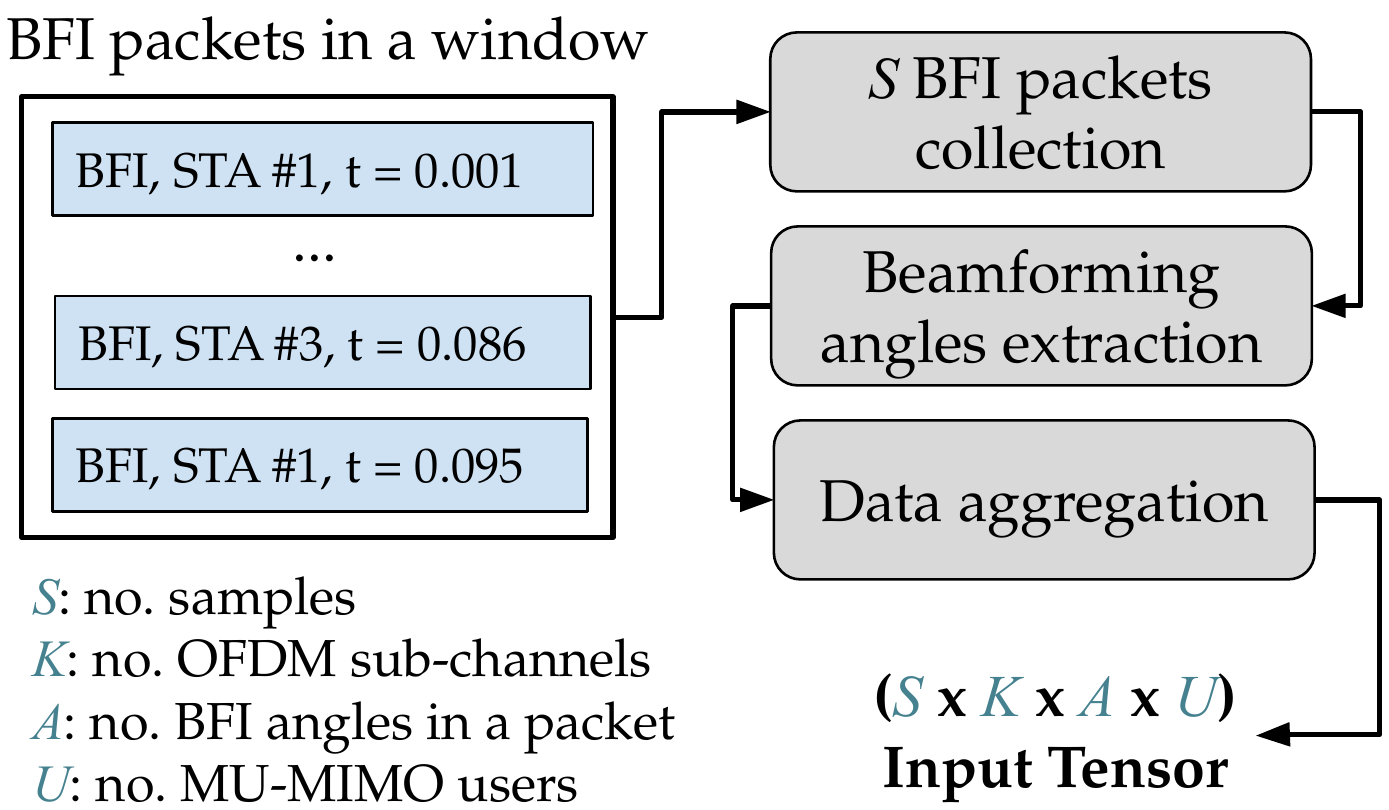}
	\caption{\bfi data processing. The processing is applied to each observation window of $W$ seconds.\vspace{-0.5cm}} 
	\label{fig:dataproc}
\end{figure}

\FW captures the \bfi reports (\textbf{step 5}), and uses the channel estimation data to perform Wi-Fi sensing. We remark that, since \mum requires fine-grained channel sounding -- every around 10 milliseconds to account for user mobility, according to \cite{gast2013802} -- it is fundamental to process the \bfi in a fast manner at the \ap. For this reason, and since cryptography would lead to excessive delays, the angles are currently sent unencrypted. Therefore, the \bfi reports are exposed to and can be read by any device that can access the wireless channel. 
Specifically, \FW relies on the \bfi transmitted by all the beamformees in the environment and captured during a time window of $W$ seconds to reliably estimate the activity being performed by a human moving within the propagation environment. This is done by using the \bfi samples collected within the window as input for a learning-based algorithm (detailed in Section~\ref{sec:learning}). Note that, as \FW leverages ongoing \gls{mum} transmissions, there is no guarantee that the same number of \bfi frames are collected within a specific time interval of $W$ seconds. This is related to the fact that we have no control on when the beamformer triggers the channel sounding procedure that generates \bfi data. Therefore, as the neural network-based classification algorithm requires the input to be of a fixed dimension, we need to determine a fixed-size input that represents the \bfi reports captured during the time window. 
The processing is applied just after having collected the data on the wireless channel (grey box in Figure~\ref{fig:beamsense-fw}) and is summarized in Figure~\ref{fig:dataproc}. Specifically, we consider the average number $S$ of \bfi packets counted (at training time) in each window during an activity recording. Windows having less than $S$ packets are padded with \bfi packets containing zero-valued angles while packets exceeding such threshold are discarded. Hence, the $K \times A$ \bfi angles contained in each packet are extracted and the final tensor is obtained by aggregating the $S \times K \times A$ angles for all the $U$ \gls{mum} users for which the \bfi data have been captured in the observation window. Note that even if it would be possible to define learning algorithms that accept input of different sizes, this would lead to an increase in the complexity of the approach, both from the training and inference perspective. Therefore, to keep the model simple for implementation on memory- and battery-constrained devices, we decided to follow a fixed-input approach.

To obtain the training data, the $S \times K \times A \times U$ tensors derived from the \bfi packets captured during the data collection phase are stored in a dataset, together with their associated activity and/or phenomenon, and a timestamp (\textbf{step 6} in Figure~\ref{fig:beamsense-fw}). This phase can be performed offline by sensing application vendors without requiring the users' cooperation. The trained model (\textbf{step 7}) is then used for online sensing (\textbf{step 8}).
As mentioned in \cite{gast2013802}, the \gls{mum} sounding procedure should be performed at least every 10~ms, which corresponds to 100 \bfi measurements/second. Since the frequency of channel sounding is not specified in the standard and since the sounding measurement lasts approximately 500 microseconds, \textit{the \bfi rate can theoretically reach 2000 \bfi per second}.\vspace{0.05cm}

\noindent\textbf{Example.}~Let assume the activity recording is 300 seconds long, and $W$ is 0.1 seconds. Then, 3000 windows are present in the recording. Let us assume that the average number of packets in the considered windows is $S$ = 10. The windows presenting less than 10 packets are zero-padded. Considering a bandwidth of 80~MHz, according to the IEEE 802.11 standard, four angles describe each of the $K$ = 234 sub-channels where sounding is performed, i.e., the total number of \gls{ofdm} sub-channels (256) minus pilots and control sub-channels that are excluded from the sounding procedure. Assuming that $U$ = 3 users are connected to the \ap, the resulting input tensor has dimensions $10 \times 234 \times 4 \times 3$, and presents a total size of $10 \cdot 234 \cdot 4 \cdot 3$ = 28080. \vspace{-0.3cm}



\subsection{The FAMReS Classification Algorithm}\label{sec:learning}

Existing research in CSI-based sensing has exposed that designing classifiers that are robust to changing the subject performing the activity (i.e., different people) and the environment where the activity is performed (i.e., different rooms) is very challenging~\cite{zheng2019zero,xiao2021onefi,meneghello2022sharp,bahadori2022rewis}. On the other hand, it is hardly feasible to collect a large amount of data for all possible scenarios. To address this key issue, we propose a deep learning (DL)-based algorithm for \bfi-based activity classification called \textit{Fast and Adaptive Micro Reptile Sensing} (FAMReS), which is a few-shot learning (FSL) algorithm based on Reptile~\cite{nichol2018first} which needs a limited set of new input data to generalize to unseen environments.

\begin{figure}[t]
	\centering
	\includegraphics[width=.38\textwidth, height=.23\textwidth]{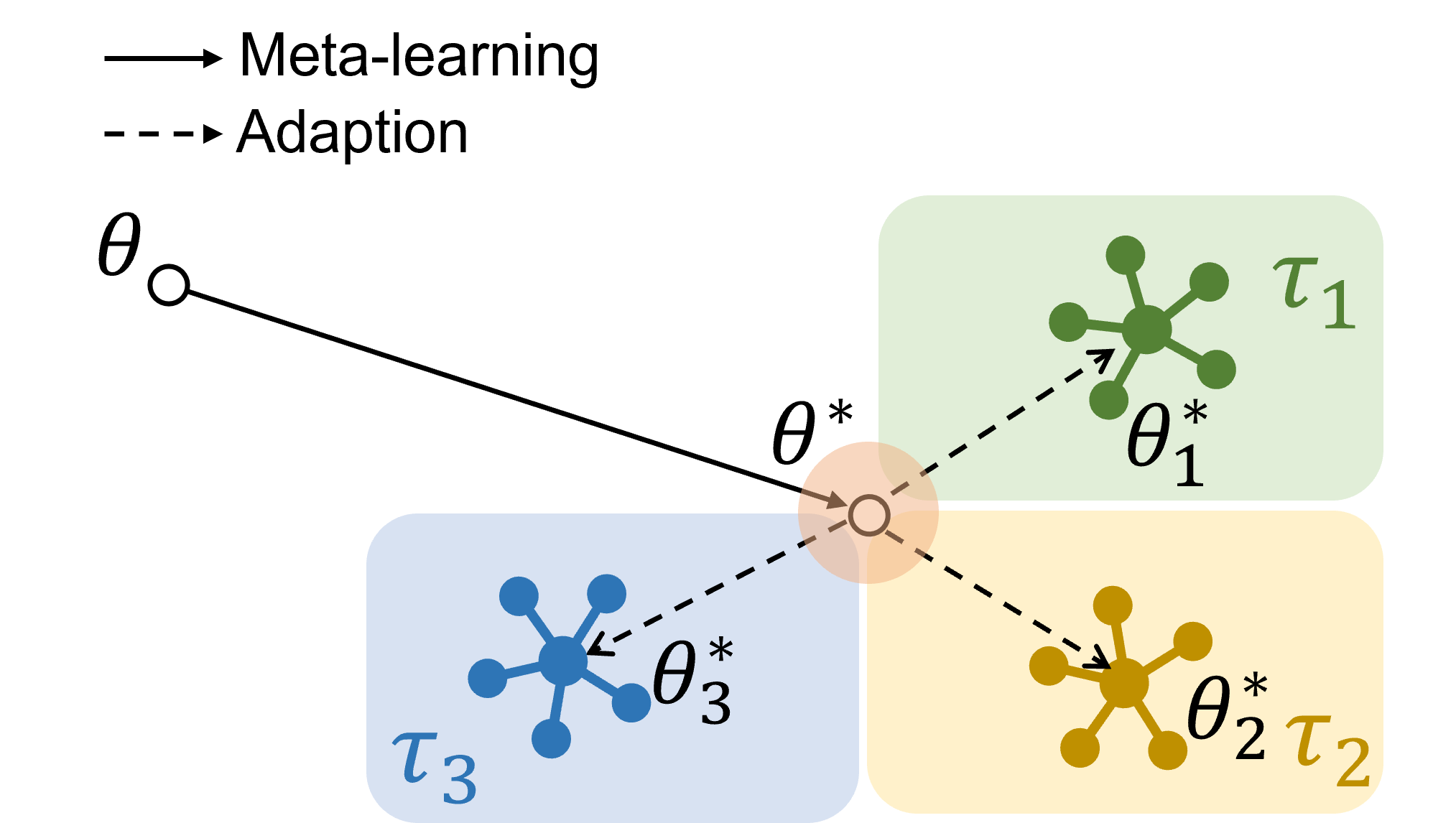}
	\caption{Example of Few-Shot Learning.\vspace{-0.6cm}}
	\label{fig:reptile}
\end{figure}

\Gls{fsl} is a DL technique that leverages only small amounts of additional data to adapt to classes that are unseen at training time. Specifically, in K-way-N-shot \gls{fsl}, the model is trained on a set of mini-batches of data that only have K different classes (ways) and N samples (shots) of each class. The key idea is that by feeding less data, the model is spurred to rapidly adapt to new tasks. This unique property makes \gls{fsl} a strong candidate to tackle the diversity of environments. \gls{fsl} can be categorized into embedding learning \cite{snell2017prototypical,vinyals2016matching}, and meta-learning \cite{nichol2018first,finn2017model}, among others. Specifically, Reptile is a gradient-based meta-learning algorithm that learns the model parameter initialization for rapid fine-tuning. The key idea is that there are some common features between different tasks that can be learned through meta-learning. Therefore, the model can be fine-tuned on a new task faster with the meta-learned weights instead of training it from the beginning. To find the initialization weights $\theta^\ast$, Reptile minimizes the expectation of the loss function $L_\tau$ with respect to the different tasks $\tau$, i.e.,

\begin{equation}
\begin{aligned}
& \theta^\ast = \underset{\theta}{\min}
& & \mathbb{E}_\tau \left\{ L_\tau\left [ f\left ( x,y |  \theta \right ) \right ] \right\},
\end{aligned}
\end{equation}
where $f (x,y|\theta)$ is the model functional approximation between input data $x$ and output $y$ obtained with parameters $\theta$.
This is equivalent to finding the $\theta^\ast$ that satisfies $ \mathbb{E}_\tau \left \{ \nabla_\theta \left( L_\tau\left [ f\left ( x,y| \theta \right ) \right ] \right) \right \} = 0$ via, e.g., stochastic gradient descent (SGD). SGD finds $\theta^\ast$ through an iterative procedure, by subsequently updating the value of $\theta$ with a new value $\theta'$ based on the gradient information:  
\begin{align}
\label{eqn:reptile1}
{\theta}' & = \theta - \beta \frac{1}{n} \sum_{\tau=1}^{n}\left ( \frac{1}{m} \sum_{i=1}^{m}\nabla_\theta \left( L_\tau\left [ f\left ( x_i,y_i| \theta \right ) \right ]\right )\right )\\
\label{eqn:reptile2}
 & = \theta - \beta \frac{1}{n} \sum_{\tau=1}^{n} \left ( \theta - \tilde{\theta} \right ),
\end{align}
where $n$ and $m$ denote the number of tasks and sampled data points of each task, respectively, $\beta$ is a scalar denoting the step size, and $\tilde{\theta}=\theta-\alpha \frac{1}{m} \sum_{i=1}^{m} \nabla_\theta \left( L_\tau\left [ f\left ( x_i, y_i| \theta \right ) \right ] \right )$ are the updated weights using $m$ sampled data from $\tau$, where $\alpha$ denotes the learning rate. $\tilde{\theta}$ can be easily obtained using any deep learning API such as TensorFlow and PyTorch. The meta-learning proceeds through the following steps: (i) sample $n$ new tasks $\{\tau\}$ with $m$ data of each task (for K-way-N-shot, $m$ is the product of K and N); (ii) compute $\tilde{\theta}$; (iii) update $\theta$ with Equation \ref{eqn:reptile2}; (iv) iterate (ii) and (iii) until the loss function stops decreasing. Figure~\ref{fig:reptile} shows how \gls{fsl} is implemented through the Reptile algorithm: once obtained the initialization weights $\theta^\ast$ through meta-learning, the model is fine-tuned on each different task.

\subsubsection{FAMReS Algorithm} \label{subsec:FAMReS Algorithm}
The original purpose of Reptile is to extract meta-features from a large dataset so that it can be quickly fine-turned when a new task is sampled from the given dataset. However, \textit{Reptile requires the inference and meta-learning data to be sampled from the same dataset}. Such a dataset should contain as many classes as possible so that the meta-learner can extract the general characteristics and fine-tune a task with fewer classes. Since this is unfeasible in \bfi-based sensing, we find some common ground between meta-learning and general DL. The aim of learning is trying to approach the ground truth between different sampled data, while meta-learning is to find shared features between various tasks. Thus, if we consider each batch of training data as a new task in meta-learning, \textit{the learning problem can be converted into a meta-learning problem}. Formally, we aim to find a set of parameters $\theta^\ast$ that minimize the loss function $L$ on training data $x_i$ and $y_i$:
\begin{equation}
\begin{aligned}
& \theta^\ast = \underset{\theta}{\min}
& & \mathbb{E}_i \left \{ L\left [ f\left ( x_i,y_i| \theta \right ) \right ] \right \}.
\end{aligned}
\end{equation}
By plugging the derivative $\mathbb{E}_i \left \{ \nabla_\theta \left (L\left [ f\left ( x_i,y_i| \theta \right ) \right ] \right ) \right \}$ to the SGD optimizer, the optimization problem can be solved as
\begin{equation}
\label{eqn:learning1}
\begin{aligned}
\tilde{\theta} & = \theta - \alpha \frac{1}{m} \sum_{i=1}^{m} \nabla_\theta \left( L\left [ f\left ( x_i,y_i| \theta \right ) \right ]\right ).\\
\end{aligned}
\end{equation}
By comparing Equation \ref{eqn:reptile1} with \ref{eqn:learning1}, we can easily find that if we set $n=1$ in Equation \ref{eqn:reptile1}, the only difference between these two equations is a constant scalar. 
Based on this observation, we note that Reptile learns common ground from different mini-batch of data. The meta-learning rate $\beta$, which is usually a scalar less than 1, is to adjust the step size of the learning, making it less likely to overfit the mini-batch data. This meta-learning process can be regarded as a warm-up phase before learning, which makes the parameters $\theta$ closer to the ground truth in the hyperspace than random initial weights. 

Inspired by this idea, FAMReS is divided into two stages: (i) meta-learning stage; and (ii) micro-learning stage. In stage (i), the model utilizes a small portion of data to learn the shared features. In stage (ii), the same micro dataset is used for training. The complete FAMReS workflow is reported in Algorithm~\ref{alg:fmrs}. 
\textbf{We stress the difference between the original Reptile and FAMReS}: we only use a small portion of data in meta-learning and micro-learning and use other unseen data for testing. On the contrary, Reptile uses the same dataset for both learning and inference. 
Although we have only done experiments offline in this work, FAMReS is a strong candidate for online learning. The algorithm can run the meta-learning phase while collecting new data. Once there is enough data, it can move on to the next stage. Therefore, we define a time variable $\delta$ in experiments to simulate the real-time implementation. We use the data collected within the $\delta$ time window for learning and the other for inference.
FAMReS is an empirical risk minimizer that can be unstable when using small values for $\delta$, depending on the distribution of training data. Meta-learning on the micro dataset can only bring the initial parameters closer to the ground truth point in the hyperspace, but the final parameters still depend on the training set. Thanks to the high stability of the \gls{bfi} data, we can always get a reasonable accuracy in the experiments unless $\delta$ is extremely small.\vspace{-0.3cm}

\RestyleAlgo{ruled}
\SetKwComment{Comment}{/*}{*/}
\SetAlgoNoLine
\LinesNotNumbered
\begin{algorithm}
\caption{The FAMReS Algorithm}\label{alg:fmrs}

Require: step size $\beta$, micro dataset $\mathbb{D} $\;
Initialize: a set of parameters $\theta$\;

\For{$iteration = 1, 2, ...$}{
sample k points of data from $\mathbb{D}$
\Comment*[r]{stage i}
compute $\tilde{\theta}$ using the SGD formulation\;
update the parameters: $\theta \leftarrow \theta + \beta \left ( \tilde{\theta} - \theta \right )$\;
}

\For{$epoch = 1, 2, ...$}{
update $\theta$ running SGD on $\mathbb{D}$\Comment*[r]{stage ii}
}
\end{algorithm}


\subsubsection{Learning Architecture} \label{subsec:learning architecture}

\begin{figure}[t]
	\centering
	\includegraphics[width=\columnwidth]{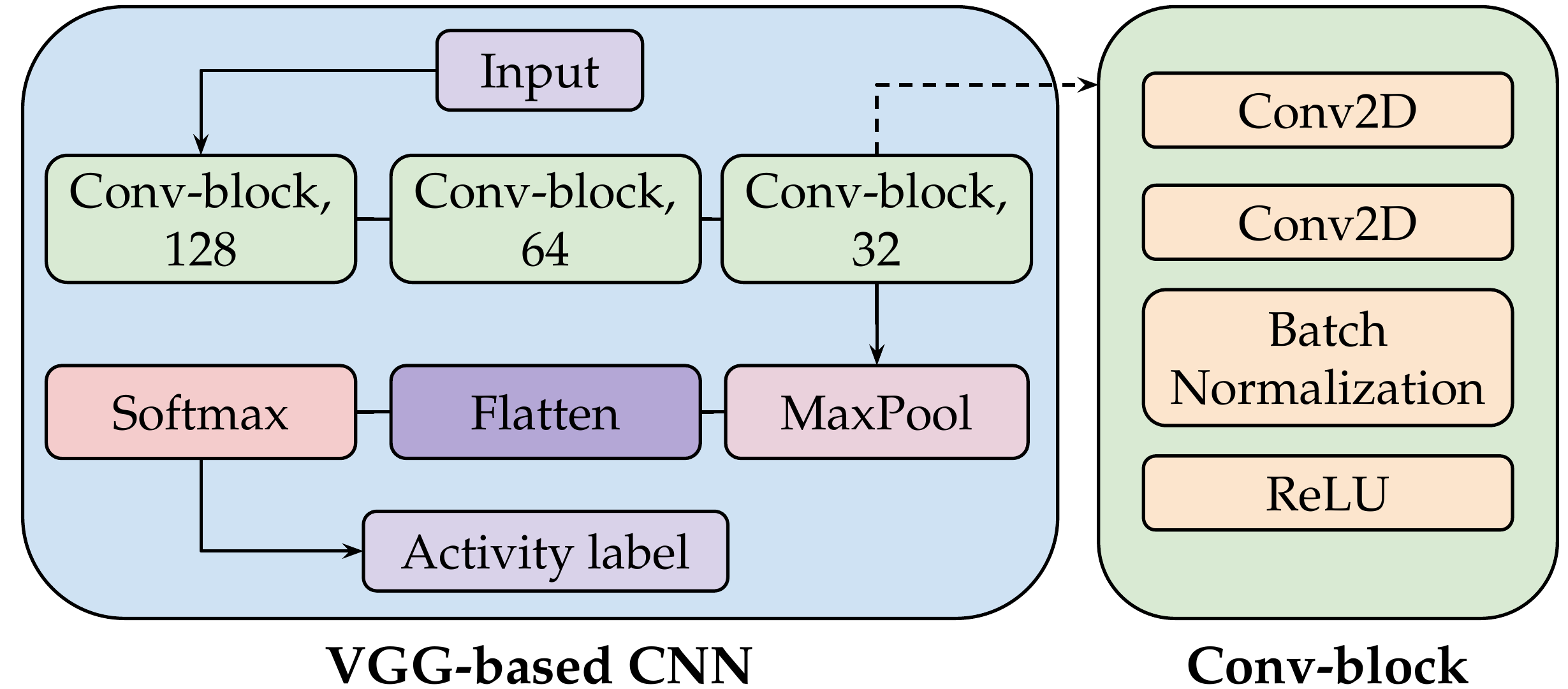}
         \setlength\abovecaptionskip{-0.3cm}
	\caption{Learning-based activity classifier.}
	\label{fig:cnn}
\end{figure}

In the last decade, \glspl{cnn} have achieved tremendous success in computer vision \cite{krizhevsky2012imagenet, simonyan2014very, he2016deep}. The convolution layer, the basis of \glspl{cnn}, can efficiently extract features by performing convolution operations on the elements of the input data. Given that in this article our aim is to investigate the effectiveness of \bfi-based sensing as compared to CSI-based sensing, we propose to use a VGG-based \cite{simonyan2014very} \gls{cnn} architecture as the human activity classifier. The network is depicted in Figure~\ref{fig:cnn} and entails stacking three convolutional blocks (\texttt{conv-block}) and a max-pooling (\texttt{MaxPool}) layer. Softmax is applied to the flattened output to obtain the probability distribution over the activity labels.

The \texttt{conv-block} is a stack of two convolution two-dimensional (2D) layers. Following the design of VGG \cite{simonyan2014very}, each convolution layer has a kernel size of $3\times3$ and a step size of $1$. To introduce non-linearity in the model, we apply a rectified linear units (ReLU) activation function at the end of each \texttt{conv-block}. Batch normalization is also used in \texttt{conv-blocks} to avoid gradient explosion or vanishing. Our VGG-based \gls{cnn} consists of three \texttt{conv-blocks} with 128, 64 and 32 filters, respectively. We choose a descending order of filters to reduce the model size since features in lower layers are usually sparser and thus require extracting more activation maps to be properly captured.

\section{Performance Evaluation}\label{sec:per_eva}

\subsection{Experimental Setup and Data Collection}\label{subsec: exp_setup}

\begin{figure}[t]
\vspace{-0.2cm}
	\centering	\includegraphics[width=\columnwidth]{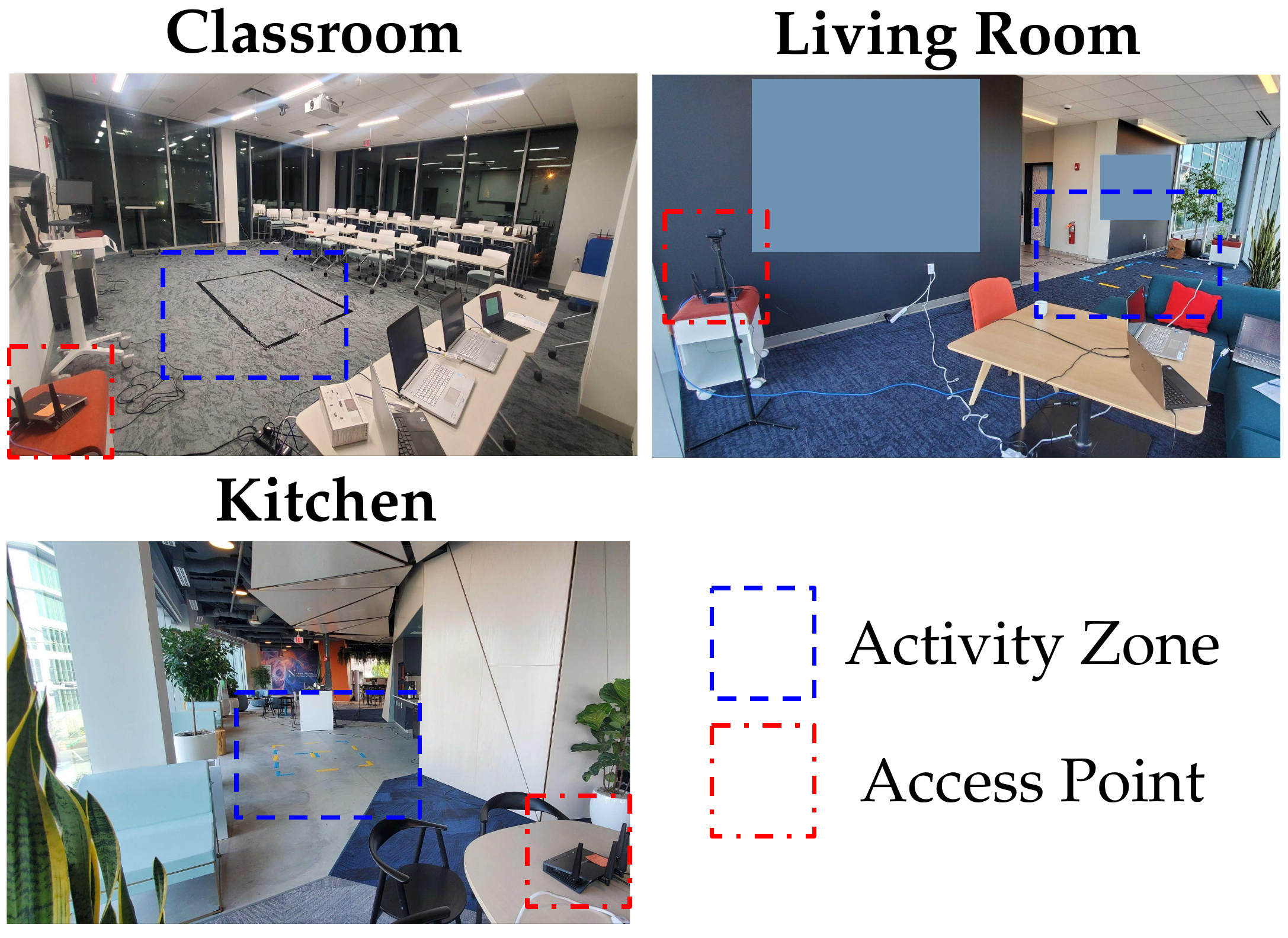} 
 \setlength\abovecaptionskip{-0.2cm}
 \caption{Sites of experimental data collection.}
\vspace{-0.4cm}
	\label{fig:pics_exp}
\end{figure}

\begin{figure}[t]
	\centering	\includegraphics[width=\columnwidth]{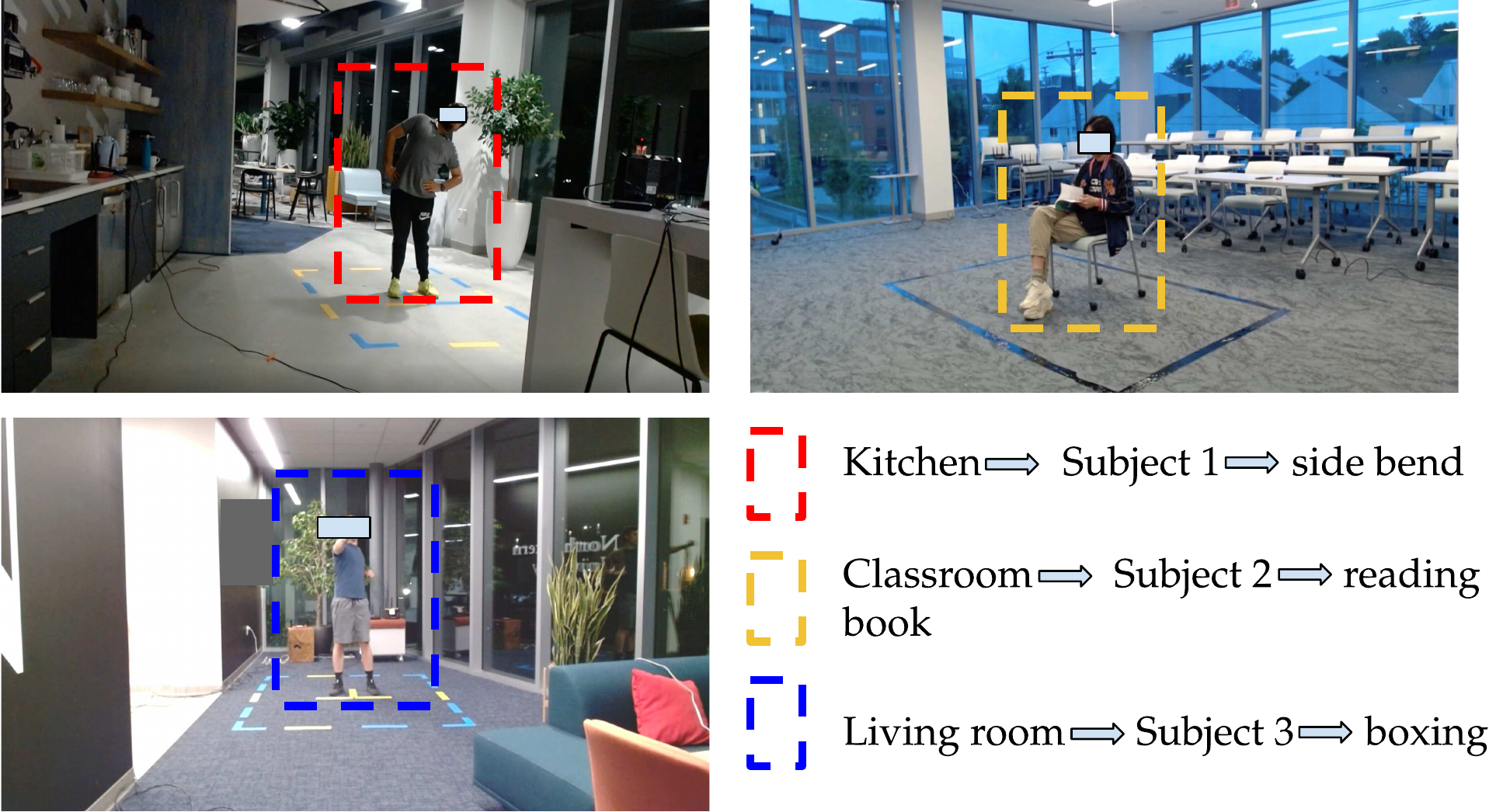} 
 \setlength\abovecaptionskip{-0.25cm}
 \caption{Sample frames from the video capture.}
 \vspace{-0.4cm}
	\label{fig:pics_sample_frame}
\end{figure}

We collected experimental data in three environments: a kitchen, a living room, and a classroom, as depicted in Figure~\ref{fig:pics_exp}. We considered three human subjects and twenty different activities: \textit{jogging, clapping, push forward, boxing, writing, brushing teeth, rotating, standing, eating, reading a book, waiving, walking, browsing phone, drinking, hands-up-down, phone call, side bend, check the wrist (watch), washing hands, and browsing laptop}. The activities are performed independently by each subject within a designated rectangular region in each of the three environments. Both \gls{bfi} and CSI data is collected for the same duration of 300 seconds for each of the twenty activities. \textbf{To create the ground truth, we captured the synchronous video streams of the subjects performing the activity}. The video streams are synchronized with the data to show what the subject is doing during the transmission of the NDP frame triggering the \bfi computation. As an example, three frames from the captured video streams are shown in Figure~\ref{fig:pics_sample_frame}.

\begin{figure}[t]
	\centering
	\includegraphics[width=\columnwidth]{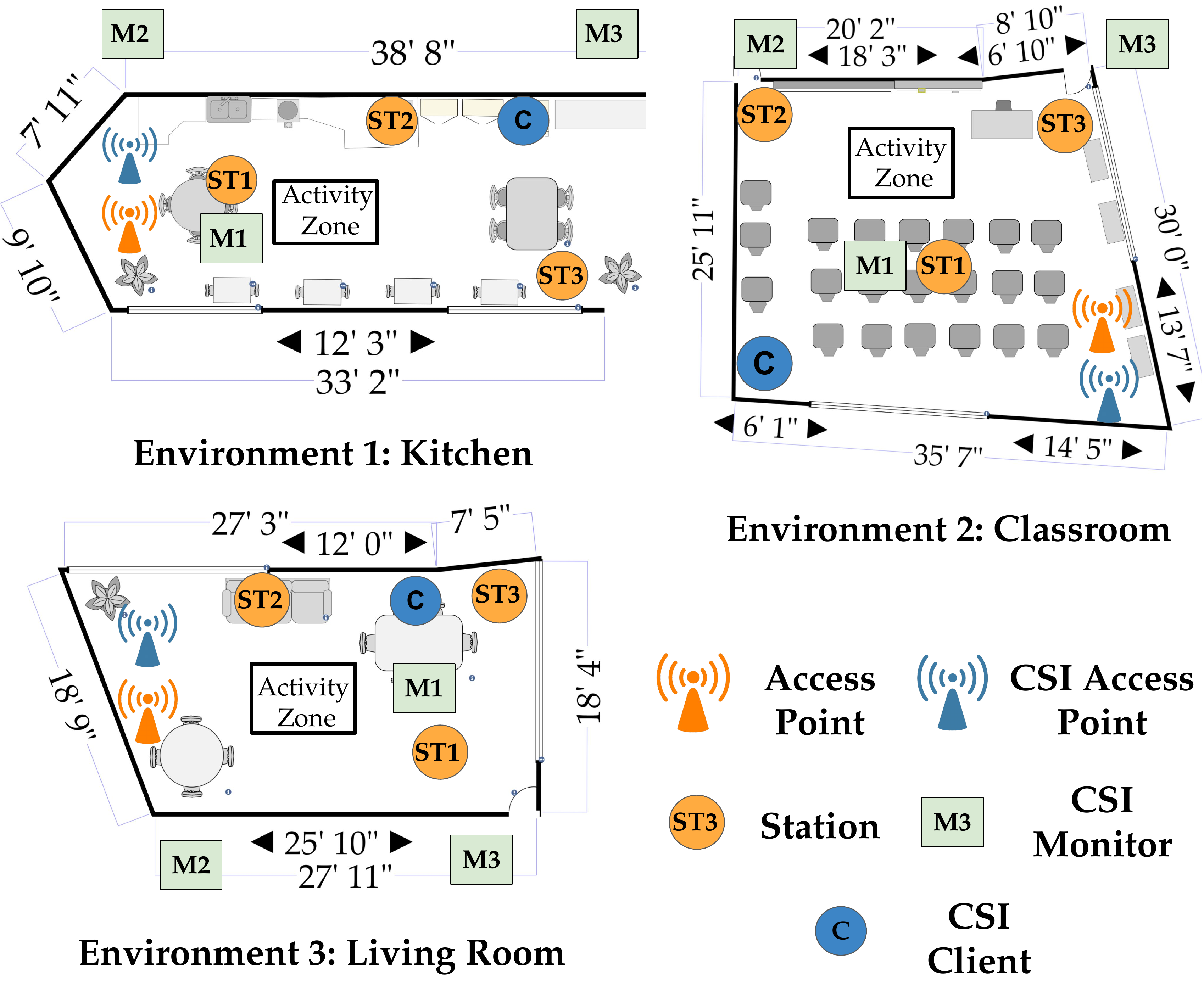} 
         \setlength\abovecaptionskip{-0.35cm}
        \caption{Experimental setups for data collection.}
	\label{fig:experimental_setup}
\end{figure}

\begin{figure}[t]
	\centering
	\subfloat [Browsing phone\label{a}]{%
		\includegraphics[width=.48\linewidth, height=.44\linewidth]{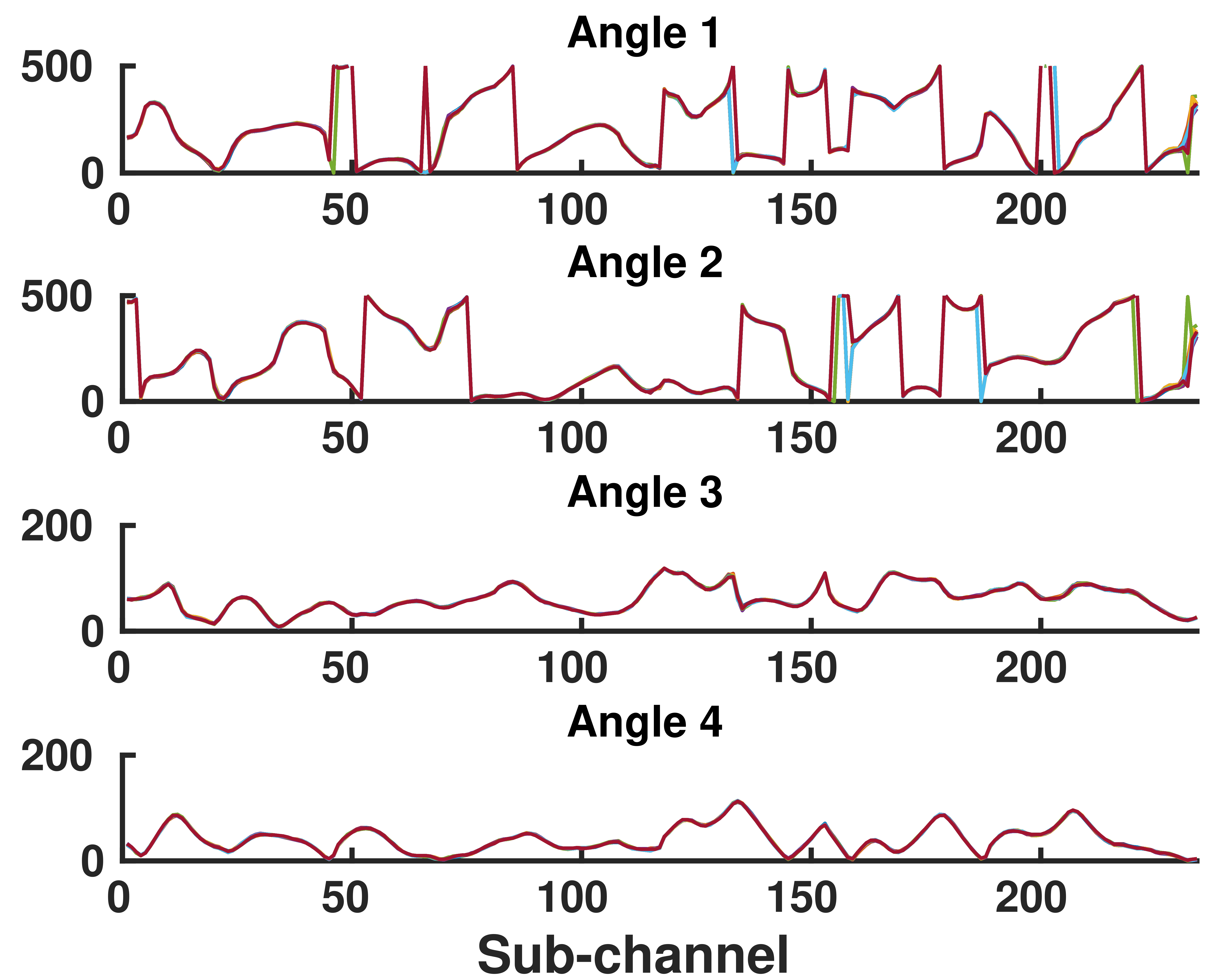}}
	\hfill
		\subfloat[Walking\label{b}]{%
		\includegraphics[width=.48\linewidth, height=.44\linewidth]{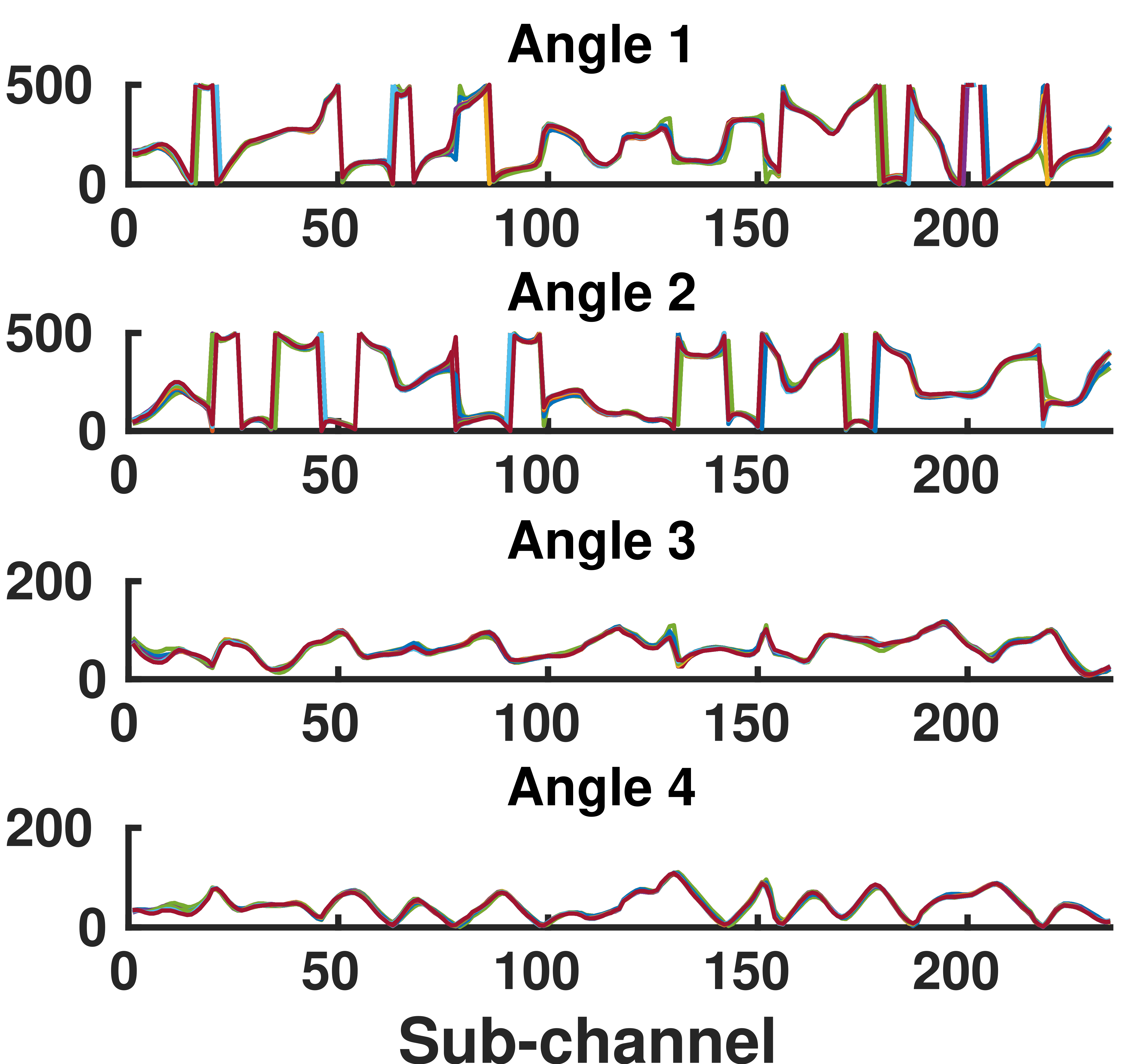}}
	\hfill
		\subfloat[Drinking water\label{c}]{%
		\includegraphics[width=.48\linewidth, height=.44\linewidth]{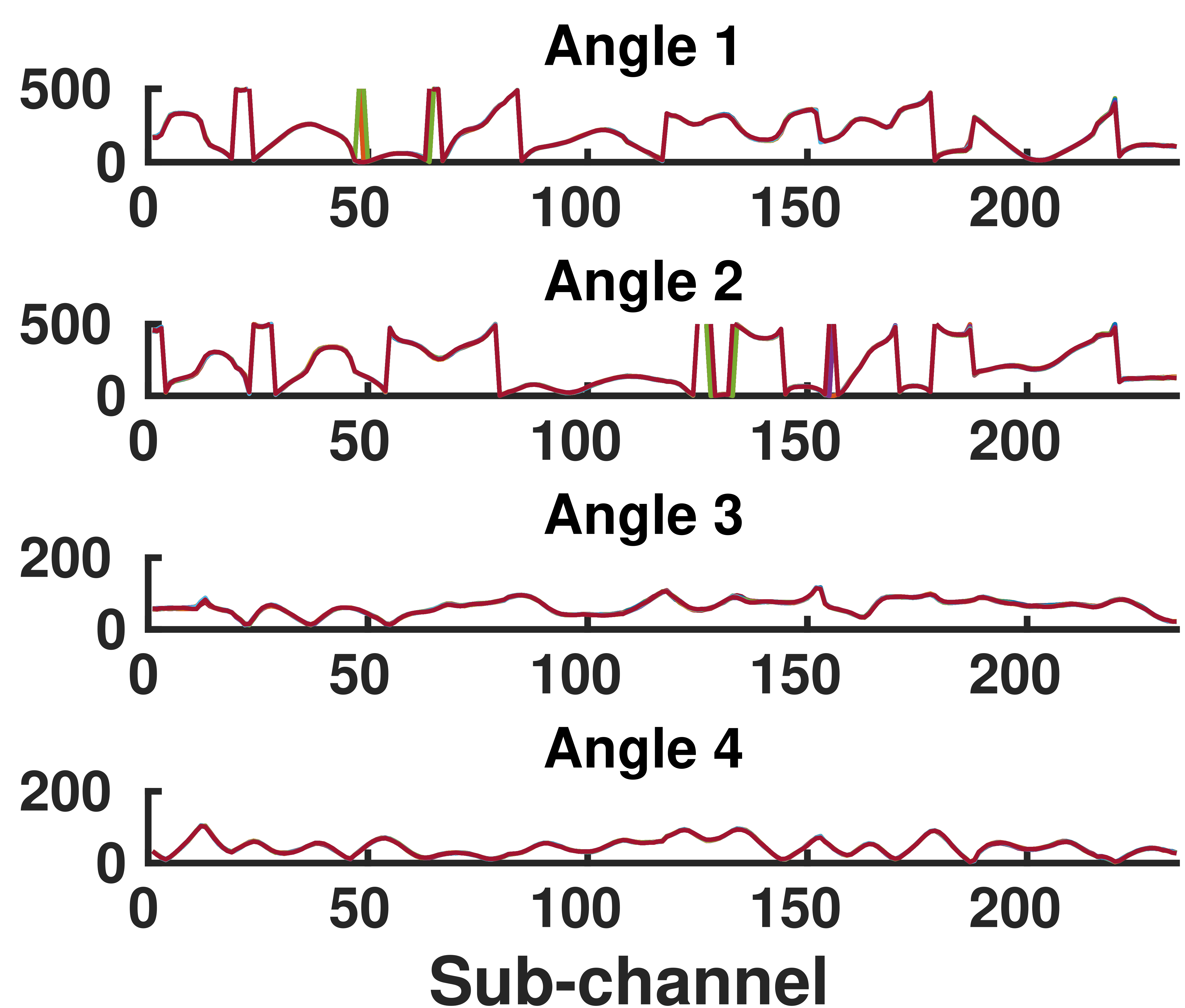}}
	\hfill
		\subfloat[Boxing\label{d}]{%
		\includegraphics[width=.48\linewidth, height=.44\linewidth]{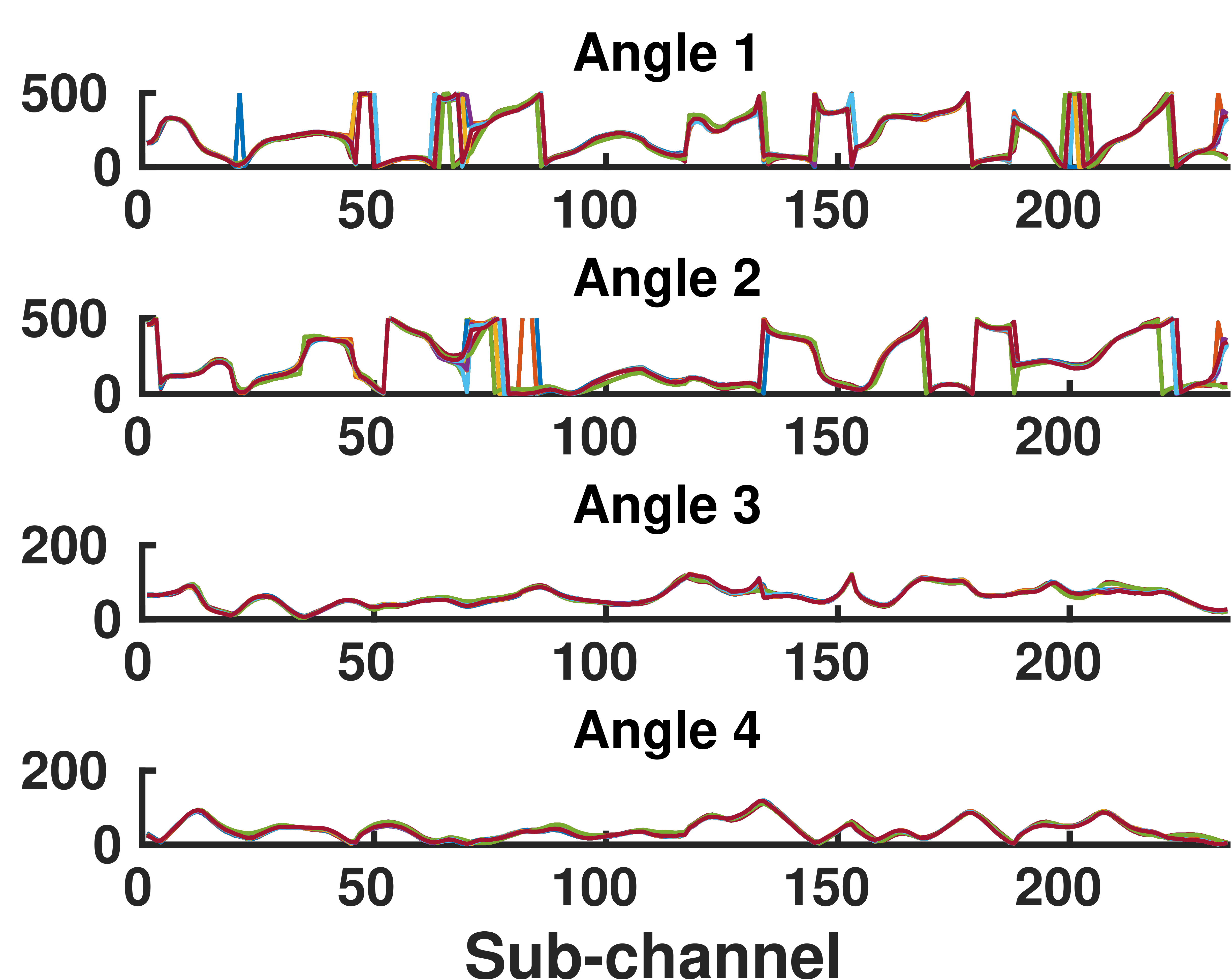}}
  
	\caption{BFI angles for each sub-channel for four activities. Each plot shows the values of 10 different packets (superimposed lines with different colors). The x-axis reports the indices of the sensed sub-channels.\vspace{-0.2cm}}
	\label{fig:angle_time} 
\end{figure}

\noindent\textbf{MU-MIMO Setup and Equipment.}~We set up an 802.11ac MU-MIMO network operating on channel 153 with center frequency $ f\textsubscript{c} $=5.77 GHz and 80 MHz bandwidth. This allows sounding $K$=234 sub-channels, i.e., 256 available sub-channels on 80 MHz channels minus 14 control sub-channels and 8 pilots. We use one AP (beamformer) and three STAs (beamformees), as depicted in Figure~\ref{fig:experimental_setup} in orange. The AP and the STAs are implemented through Netgear Nighthawk X4S AC2600 routers with $M$=3 and $N$=1 antennas enabled respectively for the AP and each of the STAs. The three STAs are served with $N\textsubscript{ss}=1$ spatial stream each and placed at three different heights and significantly spaced from each other to form a $3\times 3$ MU-MIMO system. According to the IEEE 802.11ac standard, four beamforming feedback angles (two $\phi$ and two $\psi$) are needed to represent each of the $3\times 1$ channels between the AP and the STAs. In our setup, the angle quantization process uses $b_{\phi}$= 9 bits and $b_{\psi}$ =7 bits for the feedback angles $\phi$ and $\psi$ respectively. 
UDP data streams are sent from the AP to the STAs in the downlink direction to trigger the channel sounding. The \gls{bfi} frames are captured with the Wireshark network protocol analyzer running on an off-the-shelf laptop equipped with an Intel 9560NGW wireless-AC NIC set in monitor mode. However, note that any IEEE 802.11ac-compliant NIC set in monitor mode could be used for this purpose. Moreover, notice that the frame-capturing device does not need any direct link with the AP or the STAs. The only requirement is that the capture is performed on the wireless channel where the Wi-Fi network is operating. From the captured frames, the $\phi$ and the $\psi$ angles are extracted for each of the \glspl{sta} and used as input to the \FW learning framework (see Section \ref{sec:learning}). Figure \ref{fig:angle_time} shows a sample taken from our dataset. We plot the magnitude of the four collected beamforming angles for each of the 234 available sub-channels, for ten different packets and four activities. Figure \ref{fig:angle_time} remarks that the absolute values of the angles change quite significantly among different activities, while do not change significantly among different packets. This indicates that \bfi-based sensing is a stable measurement of the channel propagation environment and thus, a strong candidate to be used within Wi-Fi sensing systems.

\noindent\textbf{CSI Network Setup and Equipment.}~For comparative studies, CSI data has also been collected concurrently with the \gls{bfi} frame capture. For this purpose, a Wi-Fi network consisting of an AP (referred to as \textit{CSI AP}) and a single \gls{sta} (referred to as \textit{CSI client}) has been set up within the same environments, as depicted in Figure~\ref{fig:experimental_setup} in blue. The network operates on the IEEE 802.11ac channel 42, i.e., the center frequency is $f_c$ = 5.21 GHz and the bandwidth is 80 MHz. The AP is implemented with a Netgear Nighthawk X4S AC2600 router, while the CSI client is a PC APU2 board equipped with an Intel 9560NGW wireless-AC NIC. For the CSI extraction, three IEEE 802.11ac-compliant Asus RT-AC86U routers (referred to as \textit{CSI monitors}) equipped with the Nexmon CSI extraction tool (\cite{nexmoncsi2019}) have been deployed, as depicted in Figure~\ref{fig:experimental_setup} in green. To have the same setup as in the MU-MIMO network, the CSI AP is enabled with $M$ = 3 antennas whereas the CSI monitors are set up to sense the channel through $N=1$ antenna over $N_{ss}$= 1 spatial stream each. UDP packets are sent from the CSI AP to the CSI client to trigger the channel estimation on the three CSI monitors. 

Note that, as shown in Figure~\ref{fig:experimental_setup}, the CSI AP and one of the CSI monitors (M1) are respectively placed at the same location as the MU-MIMO AP and one of the stations (ST1) to allow for baseline performance comparison. To show the challenges of using CSI-based sensing, we place both the BFI capturing device and the CSI monitors M2 and M3 beyond the wall of the activity zone. \textit{The CSI monitor captures the channel between itself and the CSI AP}, and, in turn, the performance decrease when CSI collectors are placed far from the monitored environment, as detailed in Section~\ref{subsec:comparison_bfi_csi}.

\subsection{Performance Analysis}

In the following, all the results are obtained with a time window size of 0.1 s with ten packets/sample with the data of three subjects combined, unless specified otherwise.

\subsubsection{Comparison between \bfi and \csi-based Sensing}\label{subsec:comparison_bfi_csi}

\begin{figure}[t]
	\centering
	\includegraphics[width=\columnwidth]{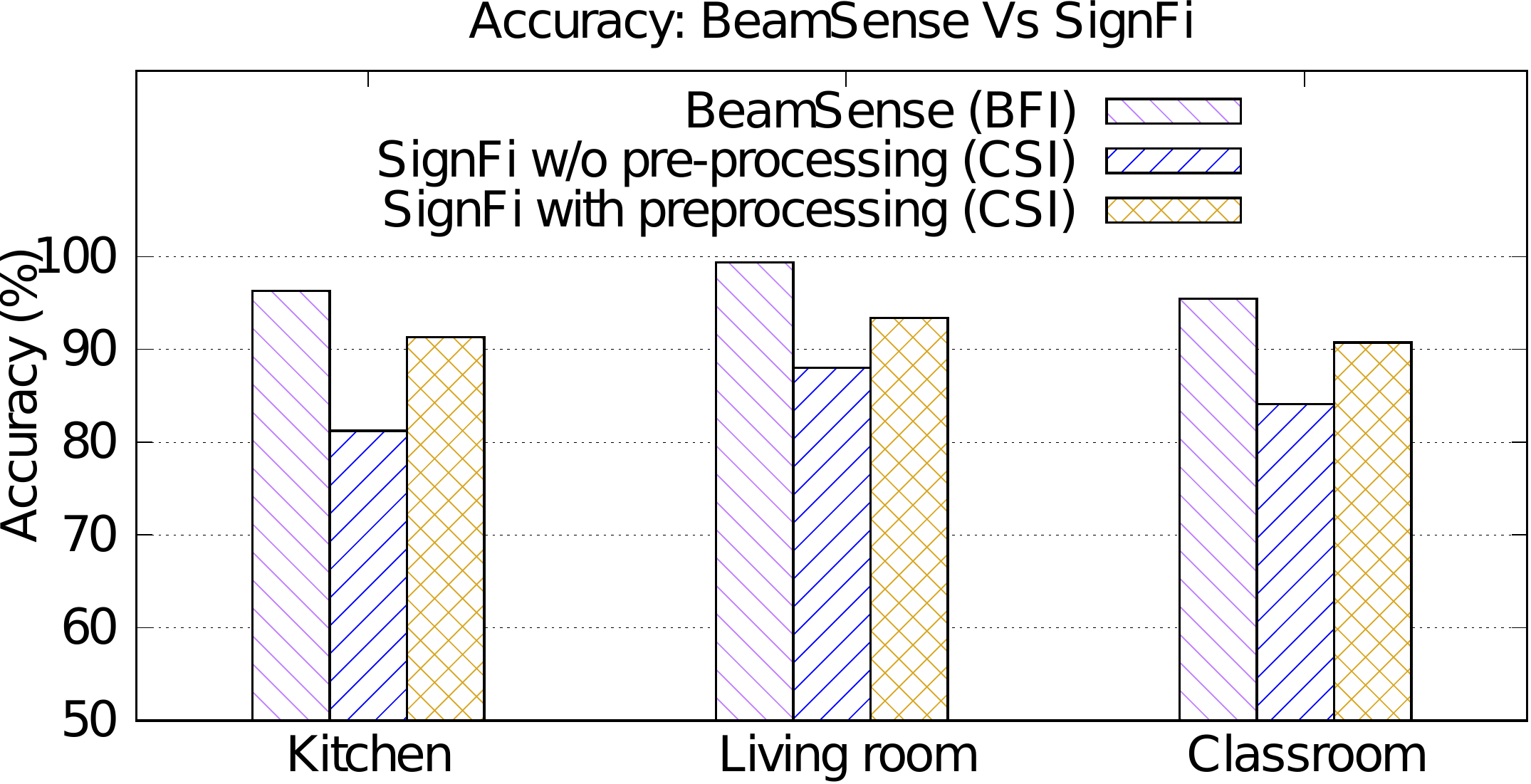}
 \setlength\abovecaptionskip{-0.2cm}
	\caption{\FW (BFI) vs SignFi (CSI) performance. \vspace{-0.1cm}} 
	\label{fig:baseline_comparison}
\end{figure}

Figure \ref{fig:baseline_comparison} shows the classification accuracy of \FW as compared to the state-of-the-art CSI-based SignFi algorithm \cite{ma2018signfi} in the three environments. For a baseline comparison, we only consider M1 and ST1 as the CSI collection device and \bfi \gls{sta} respectively which are co-located. We first evaluate the performance of \bfi and \csi-based sensing using the minimalist data processing and the \gls{cnn} architecture as referenced in Figure~\ref{fig:dataproc} and Figure~\ref{fig:cnn} respectively. The accuracy of \FW in the kitchen, living room and classroom is respectively 96\%, 99\%, and 95.47\% whereas SignFi reaches 81.19\%, 87.99\%, and 84.08\% of accuracy respectively, resulting in a 12.6\% accuracy decrease on average. We also show the performance of SignFi with the processing pipeline presented in \cite{ma2018signfi}, which unwraps the phase of each collected signal and then removes the phase noise by multiple linear regression based on the unwrapped phase across all sub-carriers and antennas. The classification accuracy improves to 91.34\%, 93\%, and 90\% in the kitchen, living room, and classroom environments, respectively. \textbf{Yet, \FW achieves better performance with no data preprocessing.} 

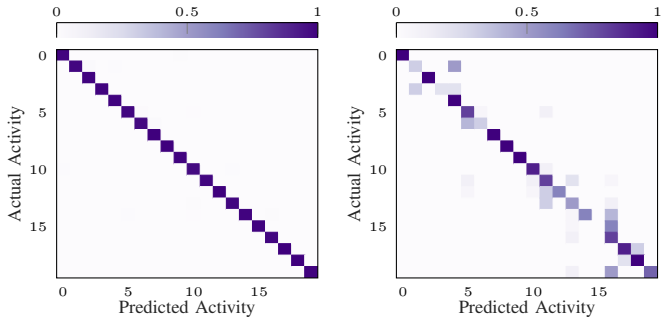
\begin{figure}[t]
    \centering
    \begin{subfigure}[t]{0.49\columnwidth}
    \centering
    \setlength\fwidth{.8\columnwidth}
    \setlength\fheight{0.7\columnwidth}
    \begin{tikzpicture}
\pgfplotsset{every tick label/.append style={font=\tiny}}

\begin{axis}[
enlargelimits=false,
colorbar,
colormap/Purples,
width=\fwidth,
height=\fheight,
at={(0\fwidth,0\fheight)},
scale only axis,
tick align=inside,
xlabel={Predicted Activity},
xmin=-0.5,
xmax=19.5,
xtick style={draw=none},
xlabel style={font=\scriptsize\color{white!15!black}},
ylabel style={font=\scriptsize\color{white!15!black}},
ylabel={Actual Activity},
ymin=-0.5,
ymax=19.5,
xlabel shift=-5pt,
ylabel shift=-5pt,
ytick style={draw=none},
axis background/.style={fill=white},
colorbar horizontal,
colorbar style={
at={(0,1.05)},               
anchor=below south west,    
width=\pgfkeysvalueof{/pgfplots/parent axis width},
xtick={0, 0.5, 1},
xmin=0,
xmax=1,
axis x line*=top,
xticklabel shift=-1pt,
point meta min=0,
point meta max=1,
},
colorbar/width=2mm,
]
\addplot [matrix plot,point meta=explicit]
 coordinates {
(0,0) [0.99] (0,1) [0.0] (0,2) [0.0015] (0,3) [0.0] (0,4) [0.0] (0,5) [0.0] (0,6) [0.0] (0,7) [0.00076] (0,8) [0.0] (0,9) [0.0] (0,10) [0.0068] (0,11) [0.0] (0,12) [0.0] (0,13) [0.0] (0,14) [0.0] (0,15) [0.0] (0,16) [0.0] (0,17) [0.0] (0,18) [0.0] (0,19) [0.0]

(1,0) [0.0] (1,1) [0.98] (1,2) [0.0014] (1,3) [0.0] (1,4) [0.0] (1,5) [0.00072] (1,6) [0.0] (1,7) [0.0] (1,8) [0.0] (1,9) [0.0] (1,10) [0.0] (1,11) [0.0] (1,12) [0.0] (1,13) [0.0] (1,14) [0.0014] (1,15) [0.0] (1,16) [0.0] (1,17) [0.0] (1,18) [0.00072] (1,19) [0.0]

(2,0) [0.0] (2,1) [0.015] (2,2) [0.99] (2,3) [0.00074] (2,4) [0.0] (2,5) [0.0015] (2,6) [0.00074] (2,7) [0.0] (2,8) [0.0] (2,9) [0.0] (2,10) [0.0] (2,11) [0.0] (2,12) [0.0] (2,13) [0.0] (2,14) [0.0] (2,15) [0.00074] (2,16) [0.0] (2,17) [0.0] (2,18) [0.0] (2,19) [0.0015]

(3,0) [0.0] (3,1) [0.0028] (3,2) [0.0028] (3,3) [0.99] (3,4) [0.0041] (3,5) [0.0] (3,6) [0.00069] (3,7) [0.0] (3,8) [0.0] (3,9) [0.0] (3,10) [0.0] (3,11) [0.0] (3,12) [0.0] (3,13) [0.00069] (3,14) [0.0] (3,15) [0.0] (3,16) [0.0] (3,17) [0.0] (3,18) [0.0] (3,19) [0.0]

(4,0) [0.0] (4,1) [0.0056] (4,2) [0.0] (4,3) [0.0] (4,4) [0.99] (4,5) [0.0007] (4,6) [0.0007] (4,7) [0.0] (4,8) [0.0] (4,9) [0.0] (4,10) [0.0007] (4,11) [0.0] (4,12) [0.0] (4,13) [0.0014] (4,14) [0.0014] (4,15) [0.0] (4,16) [0.0] (4,17) [0.0] (4,18) [0.0] (4,19) [0.0]

(5,0) [0.0] (5,1) [0.0] (5,2) [0.00072] (5,3) [0.0] (5,4) [0.0] (5,5) [0.98] (5,6) [0.0043] (5,7) [0.0] (5,8) [0.0] (5,9) [0.0] (5,10) [0.0022] (5,11) [0.00072] (5,12) [0.0022] (5,13) [0.0] (5,14) [0.0058] (5,15) [0.0] (5,16) [0.0] (5,17) [0.0] (5,18) [0.0] (5,19) [0.0]

(6,0) [0.0] (6,1) [0.00074] (6,2) [0.0015] (6,3) [0.0] (6,4) [0.00074] (6,5) [0.014] (6,6) [0.97] (6,7) [0.0] (6,8) [0.0] (6,9) [0.0] (6,10) [0.00074] (6,11) [0.0] (6,12) [0.0022] (6,13) [0.0] (6,14) [0.0] (6,15) [0.0] (6,16) [0.0] (6,17) [0.0] (6,18) [0.0] (6,19) [0.0]

(7,0) [0.0] (7,1) [0.0] (7,2) [0.0] (7,3) [0.0] (7,4) [0.0] (7,5) [0.00073] (7,6) [0.016] (7,7) [1.0] (7,8) [0.0] (7,9) [0.0] (7,10) [0.0] (7,11) [0.0] (7,12) [0.0] (7,13) [0.00073] (7,14) [0.0] (7,15) [0.0] (7,16) [0.0] (7,17) [0.0] (7,18) [0.0] (7,19) [0.0]

(8,0) [0.0] (8,1) [0.00071] (8,2) [0.0] (8,3) [0.0] (8,4) [0.0] (8,5) [0.0] (8,6) [0.0] (8,7) [0.0] (8,8) [1.0] (8,9) [0.0] (8,10) [0.0] (8,11) [0.0] (8,12) [0.0] (8,13) [0.0] (8,14) [0.0] (8,15) [0.0] (8,16) [0.0] (8,17) [0.0] (8,18) [0.0] (8,19) [0.0]

(9,0) [0.01] (9,1) [0.0] (9,2) [0.0] (9,3) [0.0] (9,4) [0.0] (9,5) [0.0] (9,6) [0.0] (9,7) [0.0] (9,8) [0.0] (9,9) [0.99] (9,10) [0.0] (9,11) [0.0] (9,12) [0.0] (9,13) [0.0] (9,14) [0.00078] (9,15) [0.0] (9,16) [0.0] (9,17) [0.0] (9,18) [0.0] (9,19) [0.0]

(10,0) [0.0] (10,1) [0.00075] (10,2) [0.00075] (10,3) [0.0] (10,4) [0.0015] (10,5) [0.0045] (10,6) [0.0015] (10,7) [0.0] (10,8) [0.0] (10,9) [0.0] (10,10) [0.97] (10,11) [0.0015] (10,12) [0.0022] (10,13) [0.0082] (10,14) [0.0045] (10,15) [0.0] (10,16) [0.0] (10,17) [0.0] (10,18) [0.00075] (10,19) [0.0]

(11,0) [0.0] (11,1) [0.0] (11,2) [0.00074] (11,3) [0.0] (11,4) [0.0] (11,5) [0.00074] (11,6) [0.0015] (11,7) [0.0] (11,8) [0.0] (11,9) [0.00074] (11,10) [0.0022] (11,11) [0.98] (11,12) [0.0022] (11,13) [0.0022] (11,14) [0.0037] (11,15) [0.0] (11,16) [0.0] (11,17) [0.0] (11,18) [0.0] (11,19) [0.0022]

(12,0) [0.0] (12,1) [0.00072] (12,2) [0.0] (12,3) [0.0] (12,4) [0.0] (12,5) [0.0] (12,6) [0.00072] (12,7) [0.0] (12,8) [0.0] (12,9) [0.0] (12,10) [0.0] (12,11) [0.0] (12,12) [1.0] (12,13) [0.0014] (12,14) [0.0] (12,15) [0.0] (12,16) [0.0] (12,17) [0.0] (12,18) [0.0] (12,19) [0.00072]

(13,0) [0.0] (13,1) [0.0] (13,2) [0.0] (13,3) [0.0] (13,4) [0.0] (13,5) [0.00072] (13,6) [0.0] (13,7) [0.0] (13,8) [0.0] (13,9) [0.0] (13,10) [0.0065] (13,11) [0.0014] (13,12) [0.0014] (13,13) [0.98] (13,14) [0.0043] (13,15) [0.0] (13,16) [0.00072] (13,17) [0.0] (13,18) [0.00072] (13,19) [0.00072]

(14,0) [0.0] (14,1) [0.0] (14,2) [0.0] (14,3) [0.0] (14,4) [0.0] (14,5) [0.00079] (14,6) [0.00079] (14,7) [0.0] (14,8) [0.0] (14,9) [0.0] (14,10) [0.00079] (14,11) [0.0016] (14,12) [0.00079] (14,13) [0.0032] (14,14) [0.99] (14,15) [0.0] (14,16) [0.0016] (14,17) [0.0] (14,18) [0.00079] (14,19) [0.0]

(15,0) [0.0] (15,1) [0.00073] (15,2) [0.0015] (15,3) [0.0] (15,4) [0.0] (15,5) [0.0] (15,6) [0.0] (15,7) [0.0] (15,8) [0.0] (15,9) [0.0] (15,10) [0.0] (15,11) [0.0015] (15,12) [0.0] (15,13) [0.0036] (15,14) [0.0036] (15,15) [0.99] (15,16) [0.0015] (15,17) [0.0] (15,18) [0.0] (15,19) [0.0015]

(16,0) [0.0] (16,1) [0.00075] (16,2) [0.0] (16,3) [0.0] (16,4) [0.0] (16,5) [0.0015] (16,6) [0.0] (16,7) [0.00075] (16,8) [0.00075] (16,9) [0.0] (16,10) [0.00075] (16,11) [0.00075] (16,12) [0.00075] (16,13) [0.0015] (16,14) [0.0089] (16,15) [0.0] (16,16) [0.98] (16,17) [0.0] (16,18) [0.0] (16,19) [0.00075]

(17,0) [0.0] (17,1) [0.0] (17,2) [0.0] (17,3) [0.0] (17,4) [0.0] (17,5) [0.0015] (17,6) [0.0] (17,7) [0.0] (17,8) [0.0] (17,9) [0.0] (17,10) [0.0] (17,11) [0.0] (17,12) [0.0] (17,13) [0.0] (17,14) [0.0] (17,15) [0.0] (17,16) [0.0] (17,17) [1.0] (17,18) [0.0015] (17,19) [0.0]

(18,0) [0.0] (18,1) [0.0] (18,2) [0.0] (18,3) [0.0] (18,4) [0.0] (18,5) [0.0015] (18,6) [0.0] (18,7) [0.0] (18,8) [0.0] (18,9) [0.0] (18,10) [0.0] (18,11) [0.0] (18,12) [0.0] (18,13) [0.0] (18,14) [0.0] (18,15) [0.0] (18,16) [0.0] (18,17) [0.0] (18,18) [1.0] (18,19) [0.0]

(19,0) [0.0] (19,1) [0.00074] (19,2) [0.00074] (19,3) [0.00074] (19,4) [0.0] (19,5) [0.0] (19,6) [0.0] (19,7) [0.0] (19,8) [0.0] (19,9) [0.0] (19,10) [0.0022] (19,11) [0.00074] (19,12) [0.0] (19,13) [0.0029] (19,14) [0.0029] (19,15) [0.0] (19,16) [0.00074] (19,17) [0.0] (19,18) [0.0] (19,19) [0.99]

};
\end{axis}
\end{tikzpicture}
    \centering
    \end{subfigure}
    \hfill
    \begin{subfigure}[t]{0.49\columnwidth}
    \centering
    \setlength\fwidth{.8\columnwidth}
    \setlength\fheight{0.7\columnwidth}
    \begin{tikzpicture}
\pgfplotsset{every tick label/.append style={font=\tiny}}

\begin{axis}[
enlargelimits=false,
colorbar,
colormap/Purples,
width=\fwidth,
height=\fheight,
at={(0\fwidth,0\fheight)},
scale only axis,
tick align=inside,
xlabel={Predicted Activity},
xmin=-0.5,
xmax=19.5,
xtick style={draw=none},
xlabel style={font=\scriptsize\color{white!15!black}},
ylabel style={font=\scriptsize\color{white!15!black}},
ylabel={Actual Activity},
ymin=-0.5,
ymax=19.5,
xlabel shift=-5pt,
ylabel shift=-5pt,
ytick style={draw=none},
axis background/.style={fill=white},
colorbar horizontal,
colorbar style={
at={(0,1.05)},               
anchor=below south west,    
width=\pgfkeysvalueof{/pgfplots/parent axis width},
xtick={0, 0.5, 1},
xmin=0,
xmax=1,
axis x line*=top,
xticklabel shift=-1pt,
point meta min=0,
point meta max=1,
},
colorbar/width=2mm,
]
\addplot [matrix plot,point meta=explicit]
 coordinates {
(0,0) [1.0] (0,1) [0.0] (0,2) [0.0] (0,3) [0.0] (0,4) [0.0] (0,5) [0.0] (0,6) [0.0] (0,7) [0.0] (0,8) [0.0] (0,9) [0.0] (0,10) [0.0] (0,11) [0.0] (0,12) [0.0] (0,13) [0.0] (0,14) [0.0] (0,15) [0.0] (0,16) [0.0] (0,17) [0.0] (0,18) [0.0] (0,19) [0.0]

(1,0) [0.0] (1,1) [0.3] (1,2) [0.0] (1,3) [0.3] (1,4) [0.0] (1,5) [0.0] (1,6) [0.0] (1,7) [0.0] (1,8) [0.0] (1,9) [0.0] (1,10) [0.0] (1,11) [0.0] (1,12) [0.0] (1,13) [0.0] (1,14) [0.0] (1,15) [0.0] (1,16) [0.0] (1,17) [0.0] (1,18) [0.0] (1,19) [0.0]

(2,0) [0.0] (2,1) [0.0] (2,2) [1.0] (2,3) [0.0] (2,4) [0.0] (2,5) [0.0] (2,6) [0.0] (2,7) [0.0] (2,8) [0.0] (2,9) [0.0] (2,10) [0.0] (2,11) [0.0] (2,12) [0.0] (2,13) [0.0] (2,14) [0.0] (2,15) [0.0] (2,16) [0.0] (2,17) [0.0] (2,18) [0.0] (2,19) [0.0]

(3,0) [0.0] (3,1) [0.0] (3,2) [0.0] (3,3) [0.2] (3,4) [0.0] (3,5) [0.0] (3,6) [0.0] (3,7) [0.0] (3,8) [0.0] (3,9) [0.0] (3,10) [0.0] (3,11) [0.0] (3,12) [0.0] (3,13) [0.0] (3,14) [0.0] (3,15) [0.0] (3,16) [0.0] (3,17) [0.0] (3,18) [0.0] (3,19) [0.0]

(4,0) [0.0] (4,1) [0.5] (4,2) [0.0] (4,3) [0.2] (4,4) [1.0] (4,5) [0.0] (4,6) [0.0] (4,7) [0.0] (4,8) [0.0] (4,9) [0.0] (4,10) [0.0] (4,11) [0.0] (4,12) [0.0] (4,13) [0.0] (4,14) [0.0] (4,15) [0.0] (4,16) [0.0] (4,17) [0.0] (4,18) [0.0] (4,19) [0.0]

(5,0) [0.0] (5,1) [0.0] (5,2) [0.0] (5,3) [0.0] (5,4) [0.0] (5,5) [0.8] (5,6) [0.4] (5,7) [0.0] (5,8) [0.0] (5,9) [0.0] (5,10) [0.0] (5,11) [0.1] (5,12) [0.05] (5,13) [0.0] (5,14) [0.0] (5,15) [0.0] (5,16) [0.0] (5,17) [0.0] (5,18) [0.0] (5,19) [0.0]

(6,0) [0.0] (6,1) [0.0] (6,2) [0.0] (6,3) [0.0] (6,4) [0.0] (6,5) [0.05] (6,6) [0.3] (6,7) [0.0] (6,8) [0.0] (6,9) [0.0] (6,10) [0.0] (6,11) [0.0] (6,12) [0.0] (6,13) [0.0] (6,14) [0.0] (6,15) [0.0] (6,16) [0.0] (6,17) [0.0] (6,18) [0.0] (6,19) [0.0]

(7,0) [0.0] (7,1) [0.0] (7,2) [0.0] (7,3) [0.0] (7,4) [0.0] (7,5) [0.0] (7,6) [0.0] (7,7) [1.0] (7,8) [0.0] (7,9) [0.0] (7,10) [0.0] (7,11) [0.0] (7,12) [0.0] (7,13) [0.0] (7,14) [0.0] (7,15) [0.0] (7,16) [0.0] (7,17) [0.0] (7,18) [0.0] (7,19) [0.0]

(8,0) [0.0] (8,1) [0.0] (8,2) [0.0] (8,3) [0.0] (8,4) [0.0] (8,5) [0.0] (8,6) [0.0] (8,7) [0.0] (8,8) [1.0] (8,9) [0.0] (8,10) [0.0] (8,11) [0.0] (8,12) [0.0] (8,13) [0.0] (8,14) [0.0] (8,15) [0.0] (8,16) [0.0] (8,17) [0.0] (8,18) [0.0] (8,19) [0.0]

(9,0) [0.0] (9,1) [0.0] (9,2) [0.0] (9,3) [0.0] (9,4) [0.0] (9,5) [0.0] (9,6) [0.0] (9,7) [0.0] (9,8) [0.0] (9,9) [1.0] (9,10) [0.0] (9,11) [0.0] (9,12) [0.0] (9,13) [0.0] (9,14) [0.0] (9,15) [0.0] (9,16) [0.0] (9,17) [0.0] (9,18) [0.0] (9,19) [0.0]

(10,0) [0.0] (10,1) [0.0] (10,2) [0.0] (10,3) [0.0] (10,4) [0.0] (10,5) [0.0] (10,6) [0.0] (10,7) [0.0] (10,8) [0.0] (10,9) [0.0] (10,10) [0.9] (10,11) [0.1] (10,12) [0.05] (10,13) [0.0] (10,14) [0.0] (10,15) [0.0] (10,16) [0.0] (10,17) [0.0] (10,18) [0.0] (10,19) [0.0]

(11,0) [0.0] (11,1) [0.0] (11,2) [0.0] (11,3) [0.0] (11,4) [0.0] (11,5) [0.1] (11,6) [0.0] (11,7) [0.0] (11,8) [0.0] (11,9) [0.0] (11,10) [0.1] (11,11) [0.8] (11,12) [0.3] (11,13) [0.3] (11,14) [0.0] (11,15) [0.0] (11,16) [0.0] (11,17) [0.0] (11,18) [0.0] (11,19) [0.0]

(12,0) [0.0] (12,1) [0.0] (12,2) [0.0] (12,3) [0.0] (12,4) [0.0] (12,5) [0.0] (12,6) [0.0] (12,7) [0.0] (12,8) [0.0] (12,9) [0.0] (12,10) [0.0] (12,11) [0.0] (12,12) [0.6] (12,13) [0.0] (12,14) [0.0] (12,15) [0.0] (12,16) [0.0] (12,17) [0.0] (12,18) [0.0] (12,19) [0.0]

(13,0) [0.0] (13,1) [0.0] (13,2) [0.0] (13,3) [0.0] (13,4) [0.0] (13,5) [0.0] (13,6) [0.0] (13,7) [0.0] (13,8) [0.0] (13,9) [0.0] (13,10) [0.0] (13,11) [0.2] (13,12) [0.0] (13,13) [0.5] (13,14) [0.05] (13,15) [0.05] (13,16) [0.1] (13,17) [0.0] (13,18) [0.0] (13,19) [0.1]

(14,0) [0.0] (14,1) [0.0] (14,2) [0.0] (14,3) [0.0] (14,4) [0.0] (14,5) [0.0] (14,6) [0.0] (14,7) [0.0] (14,8) [0.0] (14,9) [0.0] (14,10) [0.0] (14,11) [0.0] (14,12) [0.0] (14,13) [0.0] (14,14) [0.6] (14,15) [0.0] (14,16) [0.0] (14,17) [0.0] (14,18) [0.0] (14,19) [0.0]

(15,0) [0.0] (15,1) [0.0] (15,2) [0.0] (15,3) [0.0] (15,4) [0.0] (15,5) [0.0] (15,6) [0.0] (15,7) [0.0] (15,8) [0.0] (15,9) [0.0] (15,10) [0.0] (15,11) [0.0] (15,12) [0.0] (15,13) [0.0] (15,14) [0.0] (15,15) [0.0] (15,16) [0.0] (15,17) [0.0] (15,18) [0.0] (15,19) [0.0]

(16,0) [0.0] (16,1) [0.0] (16,2) [0.0] (16,3) [0.0] (16,4) [0.0] (16,5) [0.0] (16,6) [0.0] (16,7) [0.0] (16,8) [0.0] (16,9) [0.0] (16,10) [0.0] (16,11) [0.05] (16,12) [0.0] (16,13) [0.1] (16,14) [0.4] (16,15) [0.6] (16,16) [0.8] (16,17) [0.0] (16,18) [0.0] (16,19) [0.5]

(17,0) [0.0] (17,1) [0.0] (17,2) [0.0] (17,3) [0.0] (17,4) [0.0] (17,5) [0.0] (17,6) [0.0] (17,7) [0.0] (17,8) [0.0] (17,9) [0.0] (17,10) [0.0] (17,11) [0.0] (17,12) [0.0] (17,13) [0.0] (17,14) [0.0] (17,15) [0.0] (17,16) [0.0] (17,17) [0.9] (17,18) [0.2] (17,19) [0.0]

(18,0) [0.0] (18,1) [0.0] (18,2) [0.0] (18,3) [0.0] (18,4) [0.0] (18,5) [0.0] (18,6) [0.0] (18,7) [0.0] (18,8) [0.0] (18,9) [0.0] (18,10) [0.0] (18,11) [0.0] (18,12) [0.0] (18,13) [0.0] (18,14) [0.0] (18,15) [0.0] (18,16) [0.0] (18,17) [0.3] (18,18) [1.0] (18,19) [0.0]

(19,0) [0.0] (19,1) [0.0] (19,2) [0.0] (19,3) [0.0] (19,4) [0.0] (19,5) [0.0] (19,6) [0.0] (19,7) [0.0] (19,8) [0.0] (19,9) [0.0] (19,10) [0.0] (19,11) [0.0] (19,12) [0.0] (19,13) [0.0] (19,14) [0.0] (19,15) [0.0] (19,16) [0.0] (19,17) [0.0] (19,18) [0.0] (19,19) [0.7]

};
\end{axis}
\end{tikzpicture}
    \centering
    \label{fig:cm_csi}
    \end{subfigure}
     \vspace{-0.3cm}
    \setlength\abovecaptionskip{-0.2cm}
    \caption{Conf. matrices for \FW and SignFi. \vspace{-0.4cm}} 
    \label{fig:cm_feedback_BFI_csi}
\end{figure}

To shed light on which classes are the hardest to classify with CSI-based sensing, Figure~\ref{fig:cm_feedback_BFI_csi} shows the confusion matrices obtained in the kitchen using \FW and SignFi without the custom pre-processing. The bottom five classes are browsing laptop (index 20), phone call (16), hands-up-down (15), clapping (02), and boxing (04), which are indeed among the hardest classes to distinguish.

\begin{figure}[t]
	\centering
	\includegraphics[width=\columnwidth]{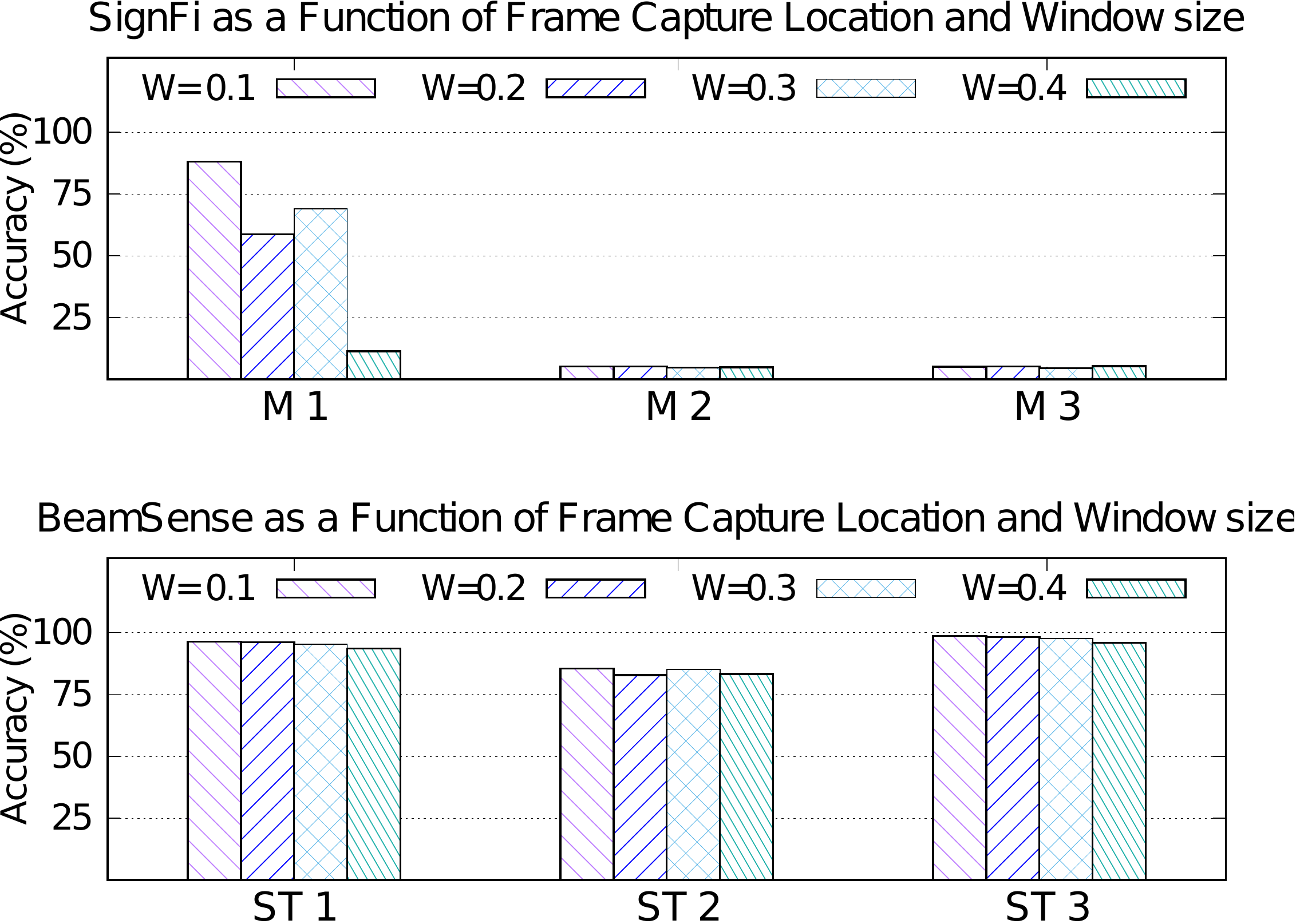}
        \setlength\abovecaptionskip{-0.3cm}
	\caption{\FW and SignFi performance changing the capture location and the window size. }
	\label{fig: capture location and window size}
\end{figure}

Figure \ref{fig: capture location and window size} shows the performance of \FW and SignFi with pre-processing evaluated in the kitchen as a function of the CSI capture location, the BFI capture location, and the window size $W$. Whereas CSI data acquired through M1 provides acceptable results since M1 is very close to the activity zone, data acquired with M2 and M3 provides poor results as M2 and M3 are far from the activity zone and beyond a wall. Specifically, the accuracy drops by 94.13\% considering an observation window size of $W$ = 0.1~s. On the contrary, the performance of \FW does not change with the location of the \bfi collector. Moreover, in the case of CSI-based sensing, we also observe a significant performance variation when varying the window size $W$. For SignFi, with the variation of the window size, the accuracy varies by 47.37\% on average, which is only 1.36\% for \FW. \textbf{This proves that feedback angles are a much more stable and reliable measurement than CSI.}

\begin{figure}[t]
	\centering
	\includegraphics[width=\columnwidth]{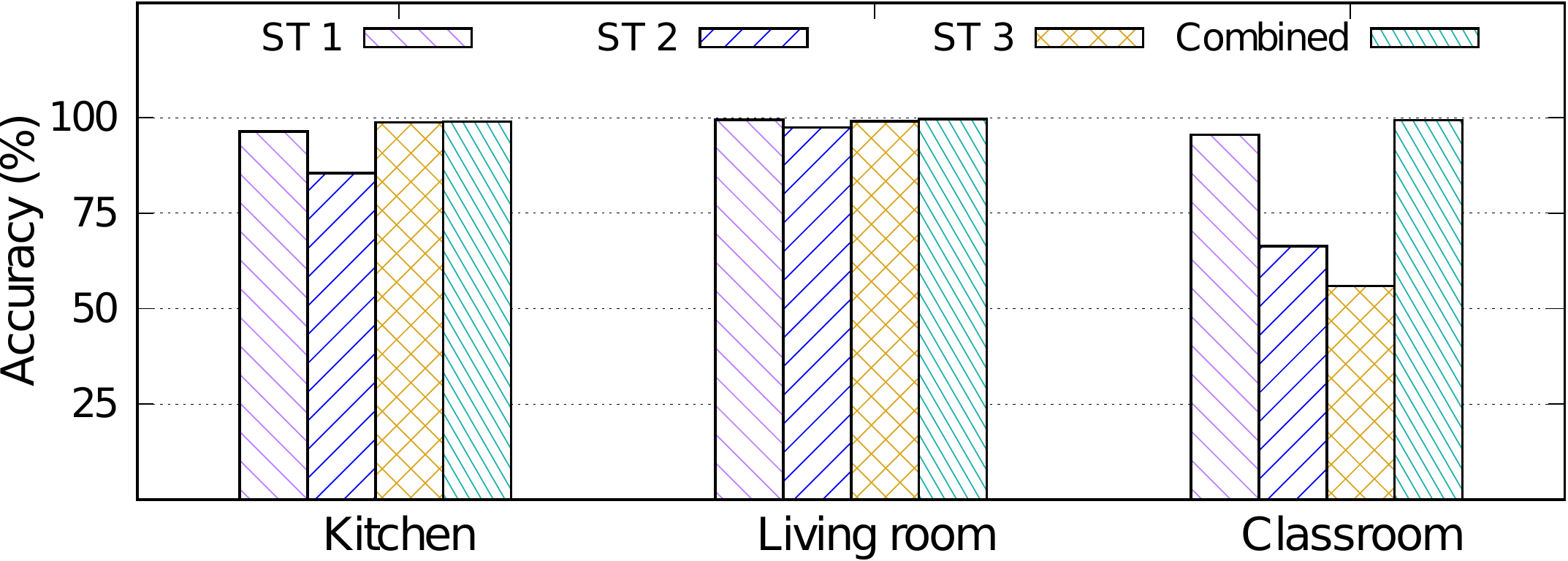}
         \setlength\abovecaptionskip{-0.3cm}
	\caption{Impact of the spatial diversity.\vspace{-0.3cm}}
	\label{fig:spatial_diversity}
\end{figure}

\subsubsection{Performance as a Function of the Spatial Diversity} \label{sub_subsection:spatial_diversity}


\begin{figure}[t]
	\centering
	\includegraphics[width=\columnwidth]{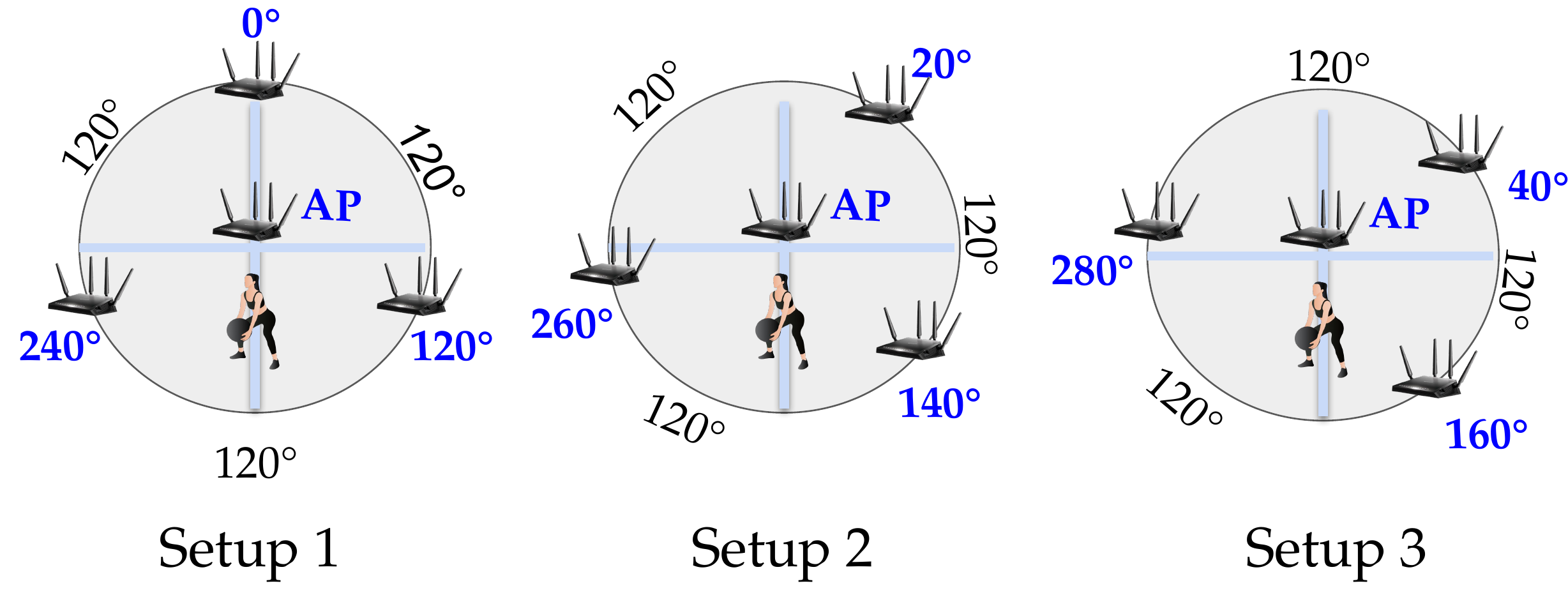}
         \setlength\abovecaptionskip{-0.4cm}
	\caption{Different setup / orientation of the STAs.}
	\label{fig:different_setup}
\end{figure}

\begin{figure}[t]
	\centering
	\includegraphics[width=\columnwidth]{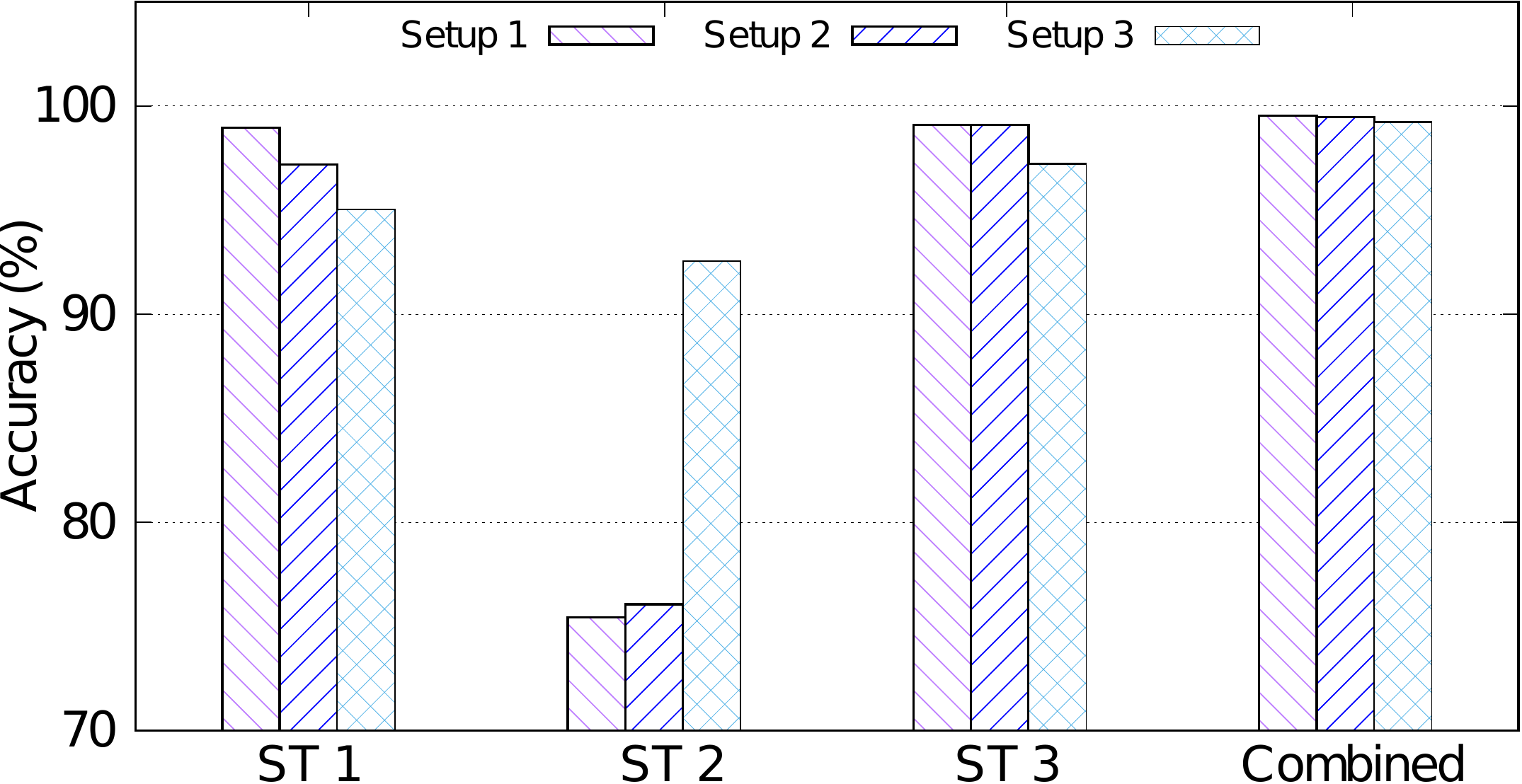}
        \setlength\abovecaptionskip{-0.3cm}
	\caption{\FW Accuracy vs STAs Location. \vspace{-0.3cm}}
	\label{fig:different_location_STAs}
\end{figure}

Figure \ref{fig:spatial_diversity} presents the performance of \FW when trained with data from a single \gls{sta} and with the combined data. First, we notice that the single \gls{sta} data is almost always a very stable measurement, with the accuracy remaining high in most of cases. However, we notice that some STAs perform worse than others, especially ST2 in the kitchen, and ST2 and ST3 in the classroom. Indeed, due to the physical location of these STAs, the communication channels between them and the AP might be in deep fade causing \FW to perform poorly. However, by aggregating the spatially diverse \gls{sta} data, \textbf{the overall accuracy is improved by up to 43.81\%} in the classroom.
Given the variability of the Wi-Fi channel, considering different \gls{sta} locations imply obtaining completely different angles for the same activity, even in the same environment, as shown in Figure~\ref{fig:spatial_diversity}. To further investigate the sensing performance as a function of the \gls{sta} location, we conduct an experiment in the kitchen entailing three different \gls{sta} locations as depicted in Figure~\ref{fig:different_setup}. The first placement is referred to as ``Setup 1'' while ``Setup 2'' and ``Setup 3'' are obtained by physically rotating each \gls{sta} by 20\textdegree clockwise, which corresponds to placing the \gls{sta} around 2 meters away from the previous location. Figure~\ref{fig:different_location_STAs} shows the accuracy of \FW in the kitchen when using data collected through each of the three setups. \FW performs very well when combining all the STAs: the accuracy is 99.53\%, 99.46\%, and 99.23\% respectively in Setup 1, Setup 2 and Setup 3. Therefore, \textbf{multi-STA sensing should be preferred over single-STA sensing whenever possible}. 

\subsubsection{Evaluation of Angle and Sub-Channel Resolution} 

It is known that Wi-Fi sensing performs worse when lowering the number of sub-channel considered in the sensing process \cite{shi2019synthesizing,zeng2020multisense}. Extensive feature extraction or higher sampling frequency can be utilized, at the cost of increasing the computational burden and intensifying pre-processing steps, as well as increasing the computational complexity of the learning process. For this reason, we investigate the trade-off between the number of angles and sub-channels considered for sensing and the sensing performance.

\begin{figure}[t]
	\centering
	\includegraphics[width=\columnwidth]{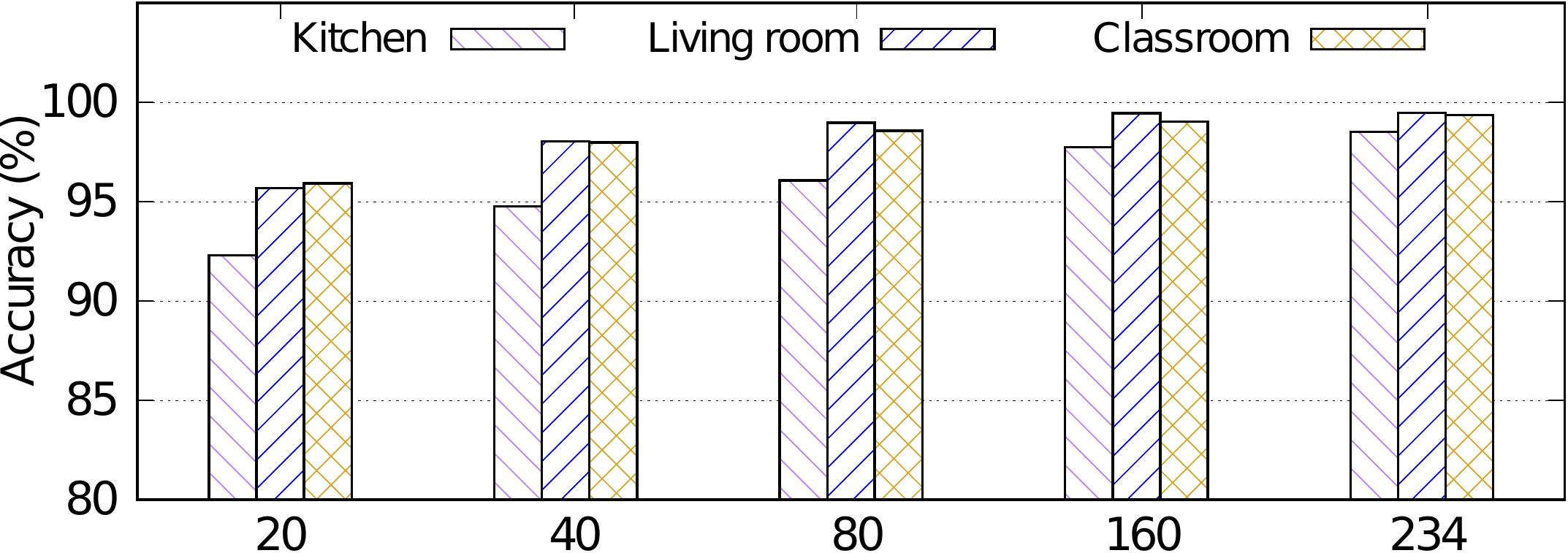}
         \setlength\abovecaptionskip{-0.3cm}
	\caption{\FW accuracy as a function of the number of sensed sub-channels.\vspace{-0.1cm}}
	\label{fig:subcarrier_resolution}
\end{figure}

Figure \ref{fig:subcarrier_resolution} shows the accuracy of \FW as a function of the number of sub-channels utilized in the learning process. To down-sample the sub-channels, we take the first 20, 40, 80, and 160 sub-channels, to emulate sensing systems with smaller available bandwidths. As expected, the accuracy decreases by 6.31\%, 3.80\%, and 3.46\% respectively for the kitchen, living room, and classroom when we switch from 234 to 20 sub-channels. However, notice that this operation drastically decreases the input tensor dimension from $10\times 234\times 12$ = 28080 to $10\times 20\times 12$ = 2400, implying that sub-channel resolution decreases the computational burden by 10$\times$ while maintaining the accuracy above 92\% in all the considered scenarios.

 \begin{figure}[t]
	\centering
	\includegraphics[width=\columnwidth]{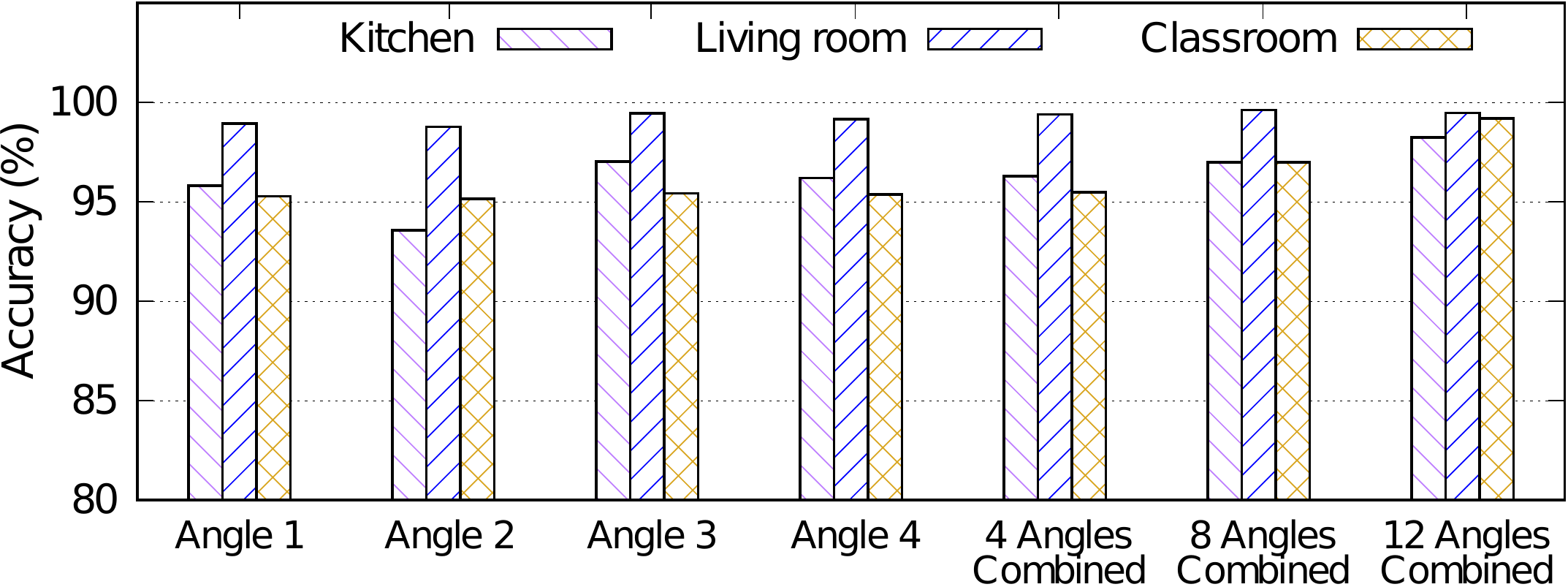}
         \setlength\abovecaptionskip{-0.3cm}
	\caption{\FW accuracy as a function of the number of the angles considered. \vspace{-0.4cm}}
	\label{fig:angle_resolution}
\end{figure}

Figure \ref{fig:angle_resolution} shows \FW performance as a function of the number of angles considered for sensing. STA1 is considered for angle 1, angle 2, angle 3, angle 4, and the combination of four angles, whereas STA1 and STA2 are considered for the combination of eight angles, and all three stations are considered for the combination of 12 angles. Figure \ref{fig:angle_resolution} shows that the accuracy decreases by 1.98\%, 0.16\%, and 2.22\% in the kitchen, living room and classroom respectively when considering a single angle with respect to the combination of 12 angles. Even though the above results show no significant variation in performance even if the angle resolution is decreased from 12 angles combined to any individual angle, we suggest aggregating at least the angles of two spatially diverse STAs to obtain a robust algorithm.

 \begin{figure}[t]
	\centering
	\includegraphics[width=\columnwidth]{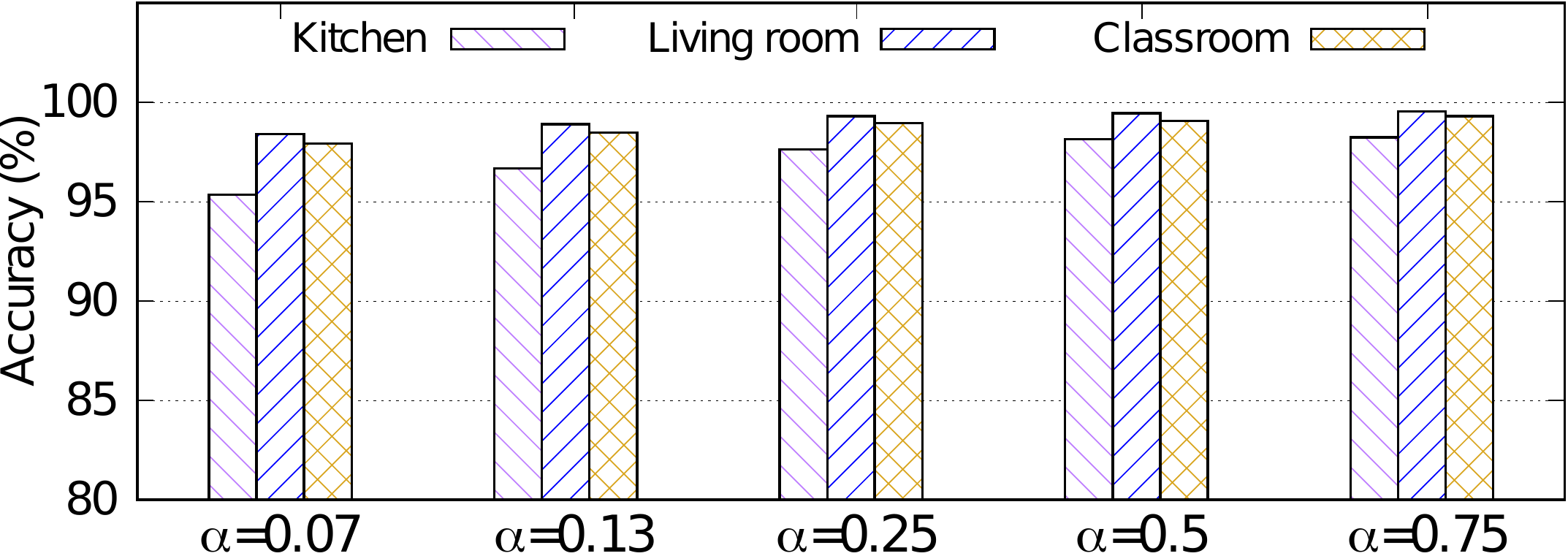}
        \setlength\abovecaptionskip{-0.3cm}
	\caption{\FW accuracy as a function of the number of the CNN filter sizes.\vspace{-0.6cm}}
	\label{fig:channelsize}
\end{figure}

\subsubsection{Evaluation of CNN Filter Size} 

To further investigate the trade-off between computation complexity and accuracy, we introduce a width multiplier $\alpha \in \left ( 0,1 \right ]$ to each layer of the \gls{cnn}-based classifier. For a given number of input channels $C$ and output channels $Z$, they become $\alpha C$ and $\alpha Z$ after applying the multiplier. Hence, the computation complexity will be reduced by $\alpha^2$ roughly. Applying the width multiplier $\alpha$ to \FW, the channel size of each \texttt{conv-block} becomes $\alpha \times 128$, $\alpha \times 64$, $\alpha \times 32$, respectively. Figure \ref{fig:channelsize} shows how the accuracy changes when applying width multiplier $\alpha \in \{ 0.07, 0.13, 0.25, 0.5, 0.75\}$. \FW accuracy, averaged over the three environments, is 97.22\%, 98.01\%, 98.62\%, 98.88\%, and 99.02\%, respectively. \textbf{As the CNN width decreases from 0.75 to 0.07, the accuracy drops marginally by 1.8\%}. This observation indicates that \FW can adapt to limited computation resources and latency-sensitive cases by sacrificing little accuracy.



\subsection{Evaluation of FAMReS Algorithm}

\begin{figure}[t]
	\centering
	\subfloat [Unseen environment\label{a}]{%
		\includegraphics[width=.48\linewidth]{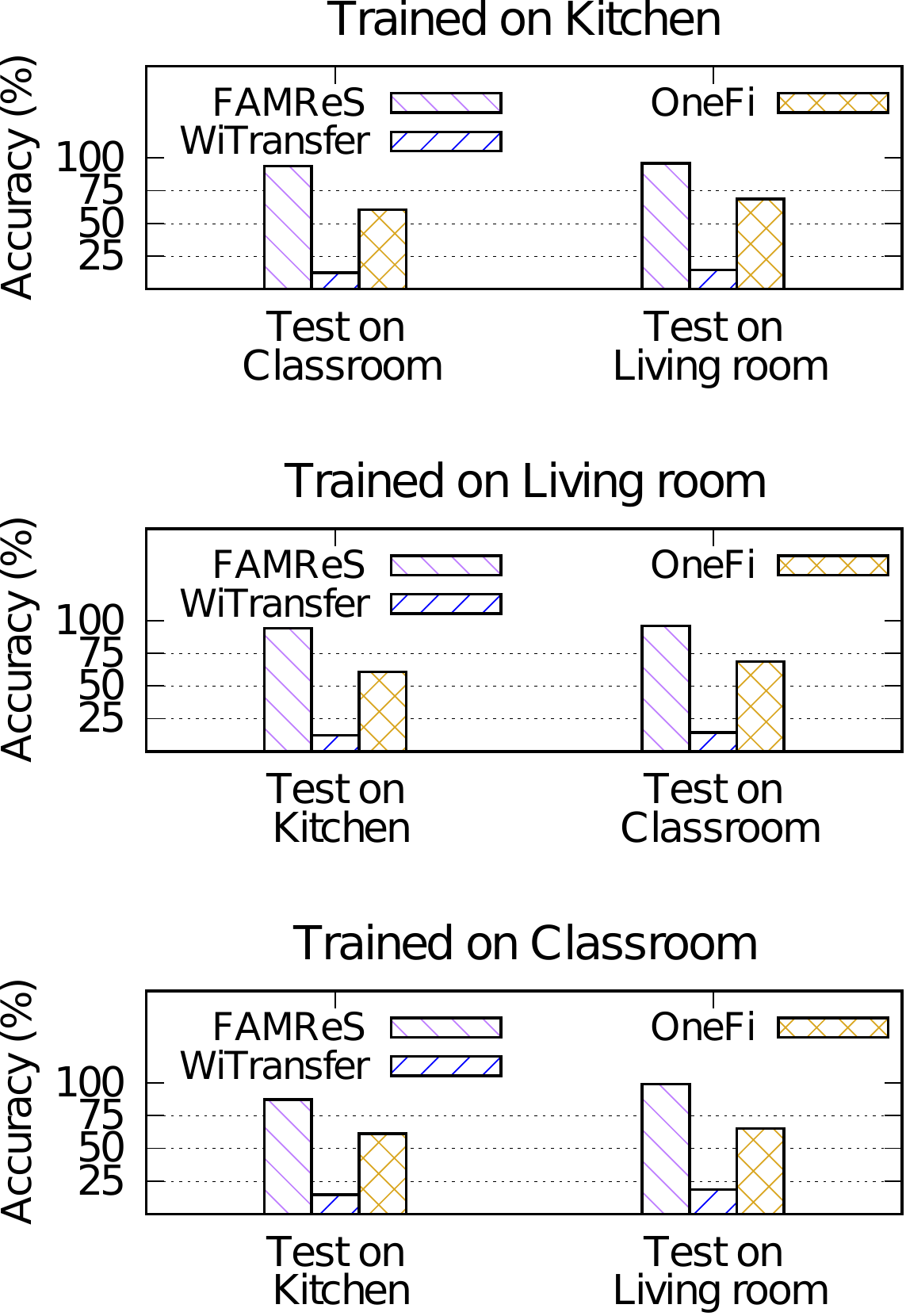}}
		\subfloat[Unseen subject\label{b}]{%
		\includegraphics[width=.48\linewidth]{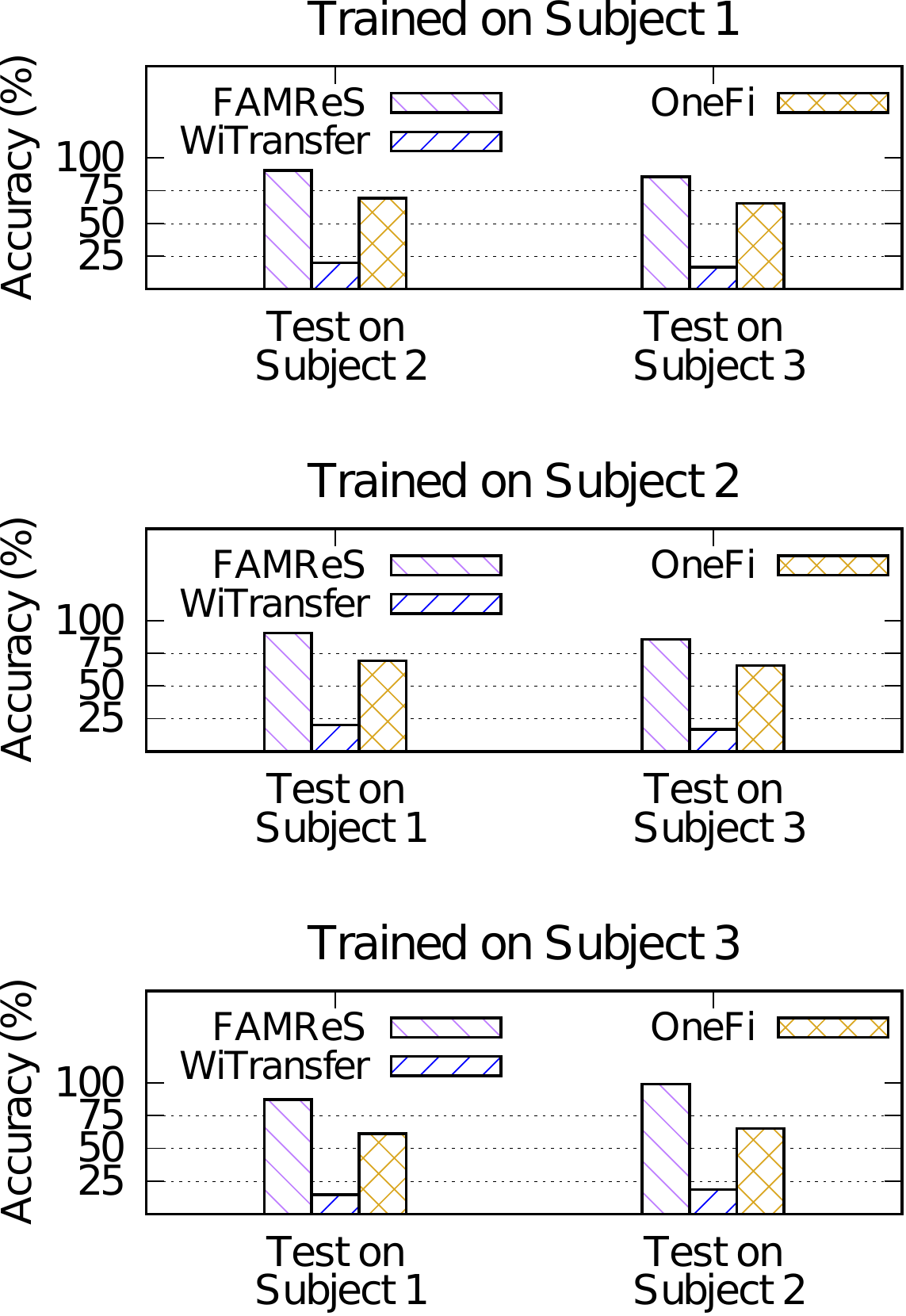}}
	\hfill
 \vspace{0.2cm}
 	\subfloat[Unseen orientation\label{b}]{%
		\includegraphics[width=.94\linewidth]{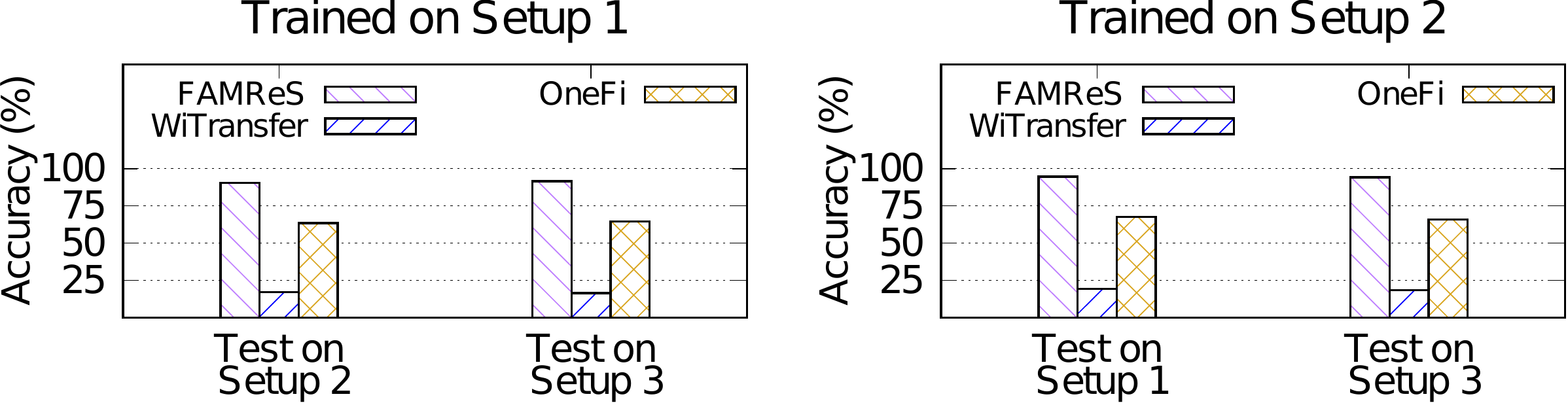}}
	\caption{Comparative analysis of \FW in unseen environments, subjects and orientations.\vspace{-0.1cm}}
	\label{fig:FAMReS_vs_CNN} 
\end{figure}

To address the challenge of generalization to unseen environments and subjects, we have proposed FAMReS in Section \ref{subsec:FAMReS Algorithm}. We compare the performance of FAMReS with the state-of-the-art \gls{fsl} algorithm OneFi\cite{xiao2021onefi} and the \gls{tl} algorithm presented in WiTransfer\cite{fang2020witransfer} for cross-domain WiFi sensing.
Figure \ref{fig:FAMReS_vs_CNN}(a) shows that with only 15~s of new data, FAMReS can adapt to new environments with an average accuracy of 94.97\%, 90.51\% and 93.09\% when trained in the kitchen, living room, and classroom respectively. On the other hand, WiTransfer achieves 13.4\%, 18.02\%, and 16.52\% respectively, which is 76.88\% less than FAMReS. The reason is that the WiTransfer pre-trained model is optimized for a specific task. Conversely, transfer learning approaches usually require more data to get rid of the data bias and 15s of new data is not enough for WiTransfer to achieve satisfactory accuracy. OneFi achieves an accuracy of 64.72\%, 63.36\%, and 63.24\% respectively in the kitchen, living room, and classroom. Although it can generalize to new environments to some extent, FAMReS performs better since it can fine-tune the whole model and learn shared information across different tasks by meta-learning. On the contrary, OneFi utilizes information from one task and only fine-tunes the classifier. Figure \ref{fig:FAMReS_vs_CNN}(b) shows a similar trend, where FAMReS is 73.41\% better than WiTransfer and 24.81\% better when compared to OneFi on average. We also evaluated the performance of FAMReS as a function of different setups as discussed in Section~\ref{sub_subsection:spatial_diversity}. Figure \ref{fig:FAMReS_vs_CNN}(c) shows that FAMReS achieves an accuracy of 90.93\%, 94.38\%, and 93.20\% when trained in setup 1, setup 2, and setup 3 respectively, and tested in the other setups. FAMReS supersedes WiTransfer and OneFi by 74.88\% and 27.28\% respectively with new unseen setups too. 

\begin{figure}[t]
	\centering
	\subfloat [Unseen environment\label{a}]{%
		\includegraphics[width=.48\linewidth]{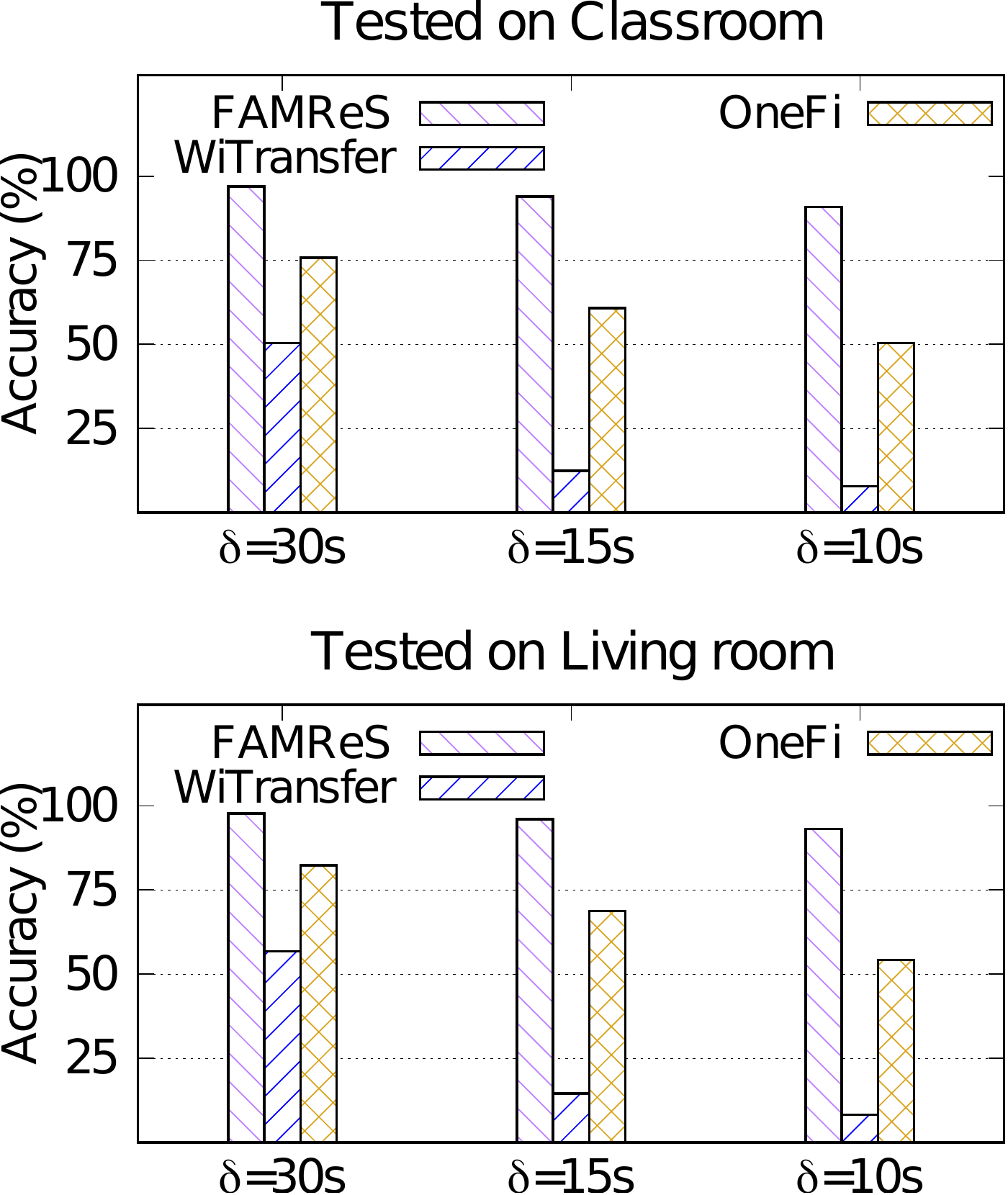}}
	\hfill
		\subfloat[Unseen subject\label{b}]{%
		\includegraphics[width=.48\linewidth]{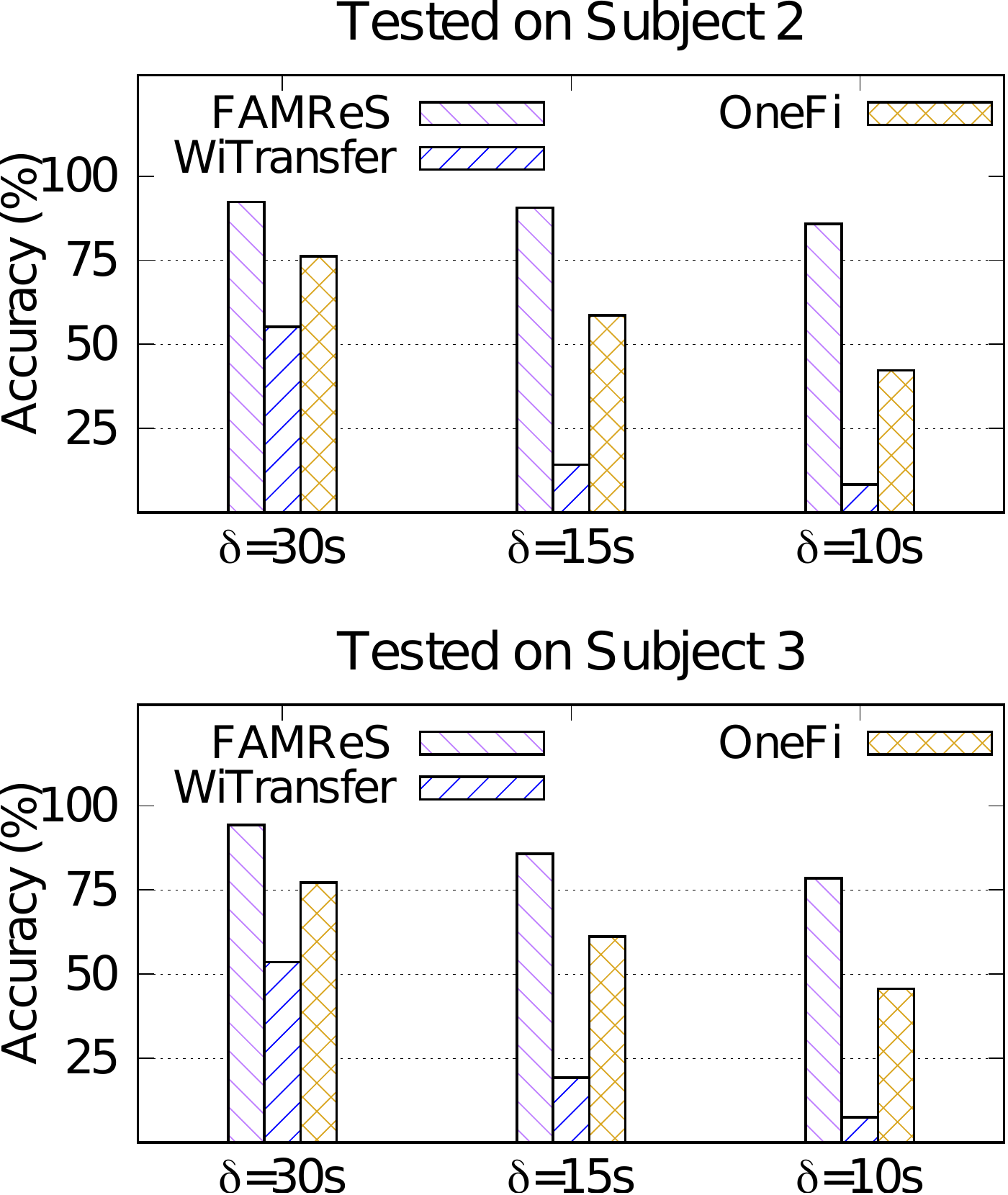}}
	\setlength\abovecaptionskip{0.2cm}
	\caption{Comparative analysis of \FW as a function of micro dataset, $\delta$.} \vspace{-0.5cm}
	\label{fig: micro_dataset_size} 
\end{figure}

Finally, we investigate the performance of FAMReS as a function of the additional micro-dataset $\delta$ required to generalize to new environments and/or subjects. Figure \ref{fig: micro_dataset_size} shows the performance of the different considered sensing algorithms as a function of the micro dataset size $\delta$. The results show that as $\delta$ decreases from 30~s to 10~s, the accuracy of FAMReS only drops by 5.30\% and 11.13\% on average when tested in unseen environments and subjects respectively. On the contrary, the performance of WiTransfer drops significantly when the duration of micro dataset $\delta$ is reduced to 10~s, showing that without the meta-learning phase, transfer learning requires more data for adaptation. Although OneFi is more stable than WiTransfer, the accuracy decreases to only 52.26\% and 43.92\% respectively with unseen environments and subjects, which is 39\% less than FAMReS. This proves the performance gain that FAMReS achieves by fine-tuning the whole network rather than fine-tuning only the classifier like OneFi.\vspace{-0.2cm}


\section{Conclusions and Remarks}\label{sec:conclusion}
In this article, we have proposed \FW, a novel approach to Wi-Fi sensing based on the usage of \mum beamforming feedback information (BFI). Conversely from CSI-based approaches, (i) the \gls{bfi} can be easily recorded by off-the-shelf devices without MIMO capabilities and without any firmware modification; (ii) the \gls{bfi} captures in a single packet the multiple channels between the \gls{ap} and the \glspl{sta}, thus achieving a much better sensing granularity. \FW includes a few-shot learning (FSL)-based classification algorithm to adapt to new environments and subjects with few additional data. We have evaluated \FW through an extensive data collection campaign involving three subjects performing twenty different activities in three indoor environments. We have compared our approach with traditional CSI-based sensing approaches and show that \FW improves the accuracy by 25\% on the average, while our FSL-based approach improves accuracy by up to 51\% when compared with state-of-the-art domain adaptive sensing models. We hope that this work will pave the way for additional research on BFI-based Wi-Fi sensing.

\bibliographystyle{IEEEtran}
\bibliography{reference}

\end{document}